\documentclass[fleqn]{article}
\usepackage{a4wide,epsfig}
\usepackage{epsfig}
\usepackage{rotating,amsmath,amssymb}

\def\nue{\ensuremath{\nu_{e}}}
\def\nubare{\ensuremath{\overline{\nu}_{e}}}

\def\numu{\ensuremath{\nu_{\mu}\ }}
\def\nubarmu{\ensuremath{\overline{\nu}_{\mu}}}

\def\nutau{\ensuremath{\nu_{\tau}\ }}

\newcommand{\numunutau}{\ensuremath{\numu \rightarrow \nutau\,}}
\newcommand{\nuenutau}{\ensuremath{\nue \rightarrow \nutau}}

\newcommand{\nubarmunubare}{\ensuremath{\overline{\nu}_\mu \rightarrow \overline{\nu}_e\,}}

\newcommand{\dmtt}{\ensuremath{\Delta m^2_{23} \,}}

\newcommand{\He}{\ensuremath{^6{\mathrm{He}\,}}}
\newcommand{\Ne}{\ensuremath{^{18}{\mathrm{Ne}\,}}}

\newcommand{\thetaot}{\ensuremath{\theta_{13}}\,}
\newcommand{\thetatt}{\ensuremath{\theta_{23}}\,}
\newcommand{\numunue}{\ensuremath{\nu_\mu \rightarrow \nu_e}}

\newcommand{\pnumunue}{\ensuremath{P(\nu_\mu \rightarrow \nu_e)}}

\newcommand{\pnubarmunubare}{\ensuremath{P(\overline{\nu}_\mu \rightarrow \overline{\nu}_e)\,}}

\newcommand{\sigdm}{\ensuremath{{\rm sign}(\Delta m^2_{23})\ }}
\newcommand{\delCP}{\ensuremath{\delta_{\rm CP}\ }}
\begin{document}

\begin{center}
{\large \bf Measurement of  three-family neutrino mixing and search for CP violation}
\vskip24pt

       {A.~Guglielmi,
 M.~Mezzetto \\{\it Istituto Nazionale di Fisica
  Nucleare, Sezione di Padova}} \\
 { P.~Migliozzi \\{\it Istituto Nazionale di Fisica
  Nucleare, Sezione di Napoli}} \\
{ F.~Terranova \\{\it Istituto Nazionale di Fisica
  Nucleare, Laboratori Nazionali di Frascati }} \\
\end{center}

\begin{abstract}
 The measurements of the parameters of the neutrino mixing matrix in the
 present and future neutrino oscillation experiments at accelerators are presented.
 The perspectives for high intensity new neutrino facilities as SuperBeams, 
 BetaBeams and Neutrino Factories devoted to precise measurements 
 subleading numu to nue oscillation parameters at the atmospheric scale
are discussed. Emphasis is on the determination of the  currently unknown
 1-3 sector of the leptonic mixing matrix (i.e.the mixing between the
first and third generation and the CP violating Dirac phase) and on
possible experimental programs to be developed in Europe.

\end{abstract}

\section{Neutrino oscillations}

The experimental evidences for neutrino
oscillations collected in the last six years represent a major
discovery in modern particle physics. The oscillation phenomenon
allows the measurement of fundamental parameters of the Standard
Model and provides the first insight beyond the electroweak scale
\cite{See-Saw}. Moreover, they
are important for many fields of astrophysics and cosmology and
open the possibility to study CP violation in the leptonic sector.

Neutrino flavor oscillations can be described in term of three mass
eigenstates $  \nu_1, \nu_2, \nu_3 $  with mass values $m_1$, $m_2$
and $m_3$ that are connected to the flavor eigenstates \nue, \numu and
\nutau by a mixing matrix $U$, usually parameterized as
\begin{equation}
 U(\theta_{12},\theta_{23},\theta_{13},\delCP)= \left(
\begin{array}{ccc}
c_{13}c_{12} & c_{13}s_{12}    &  s_{13}e^{-i\delCP} \\
-c_{23}s_{12}-s_{13}s_{23}c_{12}e^{i\delCP}&  c_{23}c_{12}-s_{13}s_{23}s_{12}e^{i\delCP}& c_{13}s_{23}      \\
 s_{23}s_{12}-s_{13}c_{23}c_{12}e^{i\delCP}     & -s_{23}c_{12}-s_{13}c_{23}s_{12}e^{i\delCP}   & c_{13}c_{13} \\
\end{array} \right)
\end{equation}

\noindent where the short-form notation $ s_{ij} \equiv \sin \theta_{ij}, c_{ij}
\equiv \cos \theta_{ij}$ is used.  As a result, the
neutrino oscillation probability depends on 3 mixing angles,
$\theta_{12},\theta_{23},\theta_{13}$, 2 mass differences, $\Delta
m^2_{12}=m_2^2-m_1^2$, $\Delta m^2_{23}=m_3^2-m_2^2$, and a CP phase $\delta_{\rm CP}$.
 Additional phases
are present in case neutrinos are Majorana particles, but they do not
influence at all neutrino flavor oscillations. Furthermore, the
neutrino mass hierarchy, the ordering with which mass eigenstates are
coupled to flavor eigenstates, can be fixed by measuring the sign of
\dmtt. In vacuum the oscillation probability between two neutrino
flavors $\alpha$, $\beta$ is:

\begin{equation}
P(\nu_\alpha \rightarrow \nu_\beta) =
 -4 \sum_{k>j}Re[W^{jk}_{\alpha\beta}] \sin^2{\frac{\Delta m^2_{jk}L}{4E_\nu}}
 \pm
 2 \sum_{k>j}Im[W^{jk}_{\alpha\beta}] \sin^2{\frac{\Delta m^2_{jk}L}{2E_\nu}}
\label{eq:osc}
\end{equation}

\noindent where $\alpha=e,\mu,\tau$, $j=1,2,3$,
$ W^{jk}_{\alpha\beta}=U_{\alpha j}U^*_{\beta j}U^*_{\alpha k}U_{\beta
k} $. 
In the case of only two neutrino flavor oscillation it can 
be written as:

\begin{equation}
     P(\nu_{\alpha} \rightarrow \nu_{\beta}) = \sin^2{2 \theta}
              {\cdot} \sin^2 \frac {1.27 \; \Delta m^2 (eV^2) {\cdot}
              L (km)}{E_{\nu} (GeV)}.
\end{equation}

\noindent Therefore two experimental parameters are relevant for neutrino
oscillations: the neutrino energy $E_\nu$ and the baseline $L$ (distance of the
neutrino  source from the detector);
 in the oscillation formula they are combined into the $L/E_\nu$ ratio.
When neutrinos pass through matter, the
oscillation probability is perturbed \cite{MSW} depending on
\sigdm \cite{Matter_Effects}.

Three claims of neutrino flavor oscillations come from
atmospheric neutrinos, solar neutrinos and beam dump experiments.

A clear evidence of \numu disappearance of atmospheric neutrinos,
through a strong zenithal modulation and an anomalous value of the
ratio of electron to muon neutrino events is reported by the
Super-Kamiokande experiment \cite{SK-Atmo}, a result well supported by
Soudan2 \cite{Soudan2} and Macro \cite{Macro} experiments.
Furthermore, Super-Kamiokande reported an oscillation signature by
directly measuring the $L/E_\nu$ parameter \cite{SK-L/E} and detected
an indirect evidence of $\nu_\tau$ appearance ruling out at 99\% C.L.
possible pure oscillations into sterile neutrinos 
\cite{SK-nutau}.
It follows that the most likely channel is an almost
pure $\numu
\rightarrow \nutau$ transition, connected with the $m_2$ and $m_3$
mass eigenstates, central values are $|\Delta m^2_{23}|=2.1{\cdot}
10^{-3}\,{\rm eV}^2$, $\sin^2 2\theta_{23}=1$ and at 90\% C.L. $1.5 {\cdot} 10^{-3}
{\rm eV}^2 < |\Delta m^2_{23}| < 3.4 {\cdot} 10^{-3}\,{\rm eV}^2$, $ \sin^2
2\theta_{23}>0.92 $ \cite{SK-Final-atm}. Mechanisms other than
neutrino oscillations,  as neutrino decay or decoherence
\cite{Pakvasa}, are almost completely ruled out by the experimental
data.

\noindent The long-baseline experiment K2K published evidence
for \numu disappearance in a neutrino beam with a mean energy $\langle
E_\nu \rangle \simeq 1.2$ GeV sent to the Super-Kamiokande detector at
a baseline of 250 km \cite{K2K}. The experimental  result:
$\sin^2{2\theta_{23}}=1.0$,
$ 1.9{\cdot} 10^{-3} < |\Delta m^2_{23}| <3.6{\cdot} 10^{-3}$ {\rm eV}$^2$
at 90 \% C.L., is in  agreement with the atmospheric data.

Solar neutrino evidence for oscillations comes from the counting
deficit of four experiments which took data at different
thresholds: Homestake \cite{Homestake}, Gallex-GNO \cite{Gallex-GNO},
 Sage \cite{Sage} and Super-Kamiokande \cite{SK-solari}. The latter experiment also
measured the energy shape distortions and day-night effects \cite{SK-day-night}. 
The spectacular comparison of the charged current,
elastic scattering and neutral current 
 rates in the SNO experiment \cite{SNO} allowed
convincing evidence of $\nu_\mu$, \nutau appearance and a first precise
determination of the oscillation parameters.

\noindent The reactor experiment KamLAND,
running at the solar $\Delta m^2$ scale, reported evidence for \nubare\
disappearance \cite{Kamland-First} in perfect agreement with the solar
data.

\noindent A combined analysis of solar plus reactor data
shows that the most likely channel is a transition of
$\nu_e$ into other active flavors ($\nu_\mu$ and $\nu_\tau$) whose
oscillation probability is modulated by $\Delta m^2_{12}$,
  central values are
$|\Delta m^2_{12}|=7.9^{+0.6}_{-0.5}{\cdot} 10^{-5}\,{\rm eV}^2$,
 $\tan^2\theta_{12}=0.40^{+0.10}_{-0.07}$
 \cite{Kamland-Final}.

A further indication of $\nubarmu \rightarrow \nubare$ oscillations
with a $\Delta m^2$ of $\,0.3 - 20 \;{\rm eV}^2$ comes from the beam dump
LSND experiment detecting a $4\sigma$ excess of \nubare\  interactions
in a neutrino beam produced by $\pi^+$ decays at rest where the \nubare\  component
 is highly suppressed ($\sim 7.8 {\cdot} 10^{-4}$) \cite{LSND}.
 The KARMEN experiment \cite{Karmen}, with a very similar technique but
with a lower sensitivity (a factor 10 less for the lower $\Delta
m^2$), and the NOMAD experiment at WANF of CERN SPS \cite{Nomad} (for
$\Delta m^2 >10$ {\rm eV}$^2$)
 do not confirm the result,  excluding a large part of the allowed region of the
 oscillation parameters.
The LSND result doesn't fit the overall picture of neutrino oscillations
and several non-standard
explanations, as for instance sterile neutrinos,
 have been put forward to solve this experimental conflict.
 The MiniBooNE experiment at FNAL,
presently taking data, is designed to settle this puzzle with
a  $5\,\sigma$ sensitivity \cite{miniboone}.

The \thetaot mixing angle represents the link between the solar and the atmospheric
neutrino oscillations: both solar and atmospheric neutrino data are compatible
with $\thetaot=0$ within the experimental sensitivity. The best experimental
constraint to \thetaot comes from the reactor experiment Chooz
\cite{Chooz}: 
$\sin^2{2\thetaot} \leq 0.14$ at 90 \% C.L. for $\dmtt=2.5 {\cdot} 10^{-3}$ eV$^2$.

The measurement of the mixing angle parameters can be performed in long-baseline
oscillation experiments with suitable neutrino beams produced at accelerators,
since this technique offers a better  control over the neutrino flux.
In particular three parameters, which are still to be measured,
the mixing angle \thetaot, the CP phase \delCP and \sigdm can be determined
by detecting sub-leading \numunue\  oscillations, as discussed in Section
\ref{section:numunue}.

\section{Present generation of long-baseline experiments}

 Over the next five years the present generation of oscillation experiments at
 accelerators with long-baseline $\nu_{\mu}$ beams (Table~\ref{tab:beams1}),
 K2K at KEK \cite{K2K}, MINOS \cite{Minos}  at the NUMI beam from FNAL \cite{NUMI} and ICARUS \cite{ICARUS} and OPERA
 \cite{OPERA} at the CNGS beam from CERN \cite{CNGS} are expected
 to confirm the
 atmospheric evidence of oscillations and measure $\sin^2 2 \theta_{23}$ and
 $|\Delta m^2_{23}|$ within $10 \div 15 $ \%  of accuracy if
 $|\Delta m^2_{23}| >  10^{-3}$ eV$^2$.
 K2K and MINOS are looking for neutrino disappearance, by measuring
 the $\nu_{\mu}$ survival probability as a function of neutrino energy while
 ICARUS and OPERA will search for evidence
 of $\nu_{\tau}$ interactions in a  $\nu_{\mu}$ beam, the final
 proof of \numunutau oscillations.
 K2K has already completed its data taking at the end of 2004, while
 MINOS has started data taking beginning 2005. CNGS is expected to start
 operations in the second half of 2006.

 \begin{table*}[htb]
 \caption{\sf Main parameters for present long-baseline neutrino beams}
 \begin{tabular}{lccccc}
 \hline
  Neutrino facility  &Proton momentum (GeV/c)&L (km)& $E_{\nu}$
 (GeV) & pot/yr ($10^{19}$)\\
 \hline
              KEK PS       &   12    & 250    &   1.5      &  2   \\
              FNAL NUMI    &  120    & 735    &   3        &  20$\div$ 34\\
              CERN CNGS    &  400    & 732    &   17.4     &  4.5$\div$ 7.6\\
 \hline
 \end{tabular}
 \label{tab:beams1}
 \end{table*}

 \noindent In all these facilities conventional muon neutrino beams are produced through the
 decay of $\pi$ and K mesons generated by a high energy proton beam hitting needle-shaped
 light targets. Positive (negative) mesons are sign-selected and focused (defocused)
 by large acceptance  magnetic lenses into a long evacuated decay
 tunnel where $\nu_{\mu}$'s  ($\overline{\nu}_{\mu}$'s) are generated.
 In case of positive charge selection, the $\nu_{\mu}$ beam has typically
 a contamination of $\overline{\nu}_{\mu}$ at few percent level
 (from the decay of the residual $\pi^{-}, K^{-}$ and $K^0$) and
 $\sim 1 \%$ of  $\nu_e$ and $\overline{\nu}_e$
 coming from three-body $K^{\pm}$, $K_0$ decays and $\mu$ decays.
 The precision on the evaluation of the intrinsic $\nu_e$ to $\nu_\mu$ contamination  is
 limited by the knowledge of the $\pi$ and $K$ production in the primary proton beam target.
 Hadroproduction measurements at 400 and 450 GeV/c performed with the NA20 \cite{NA20} and SPY
 \cite{SPY}
 experiments at the CERN SPS provided results with  $5 \div 7 \%$ intrinsic systematic uncertainties.

 \begin{figure*}[!htp]
  \begin{center}
  \vskip -0.5cm
  \mbox{\epsfig{file=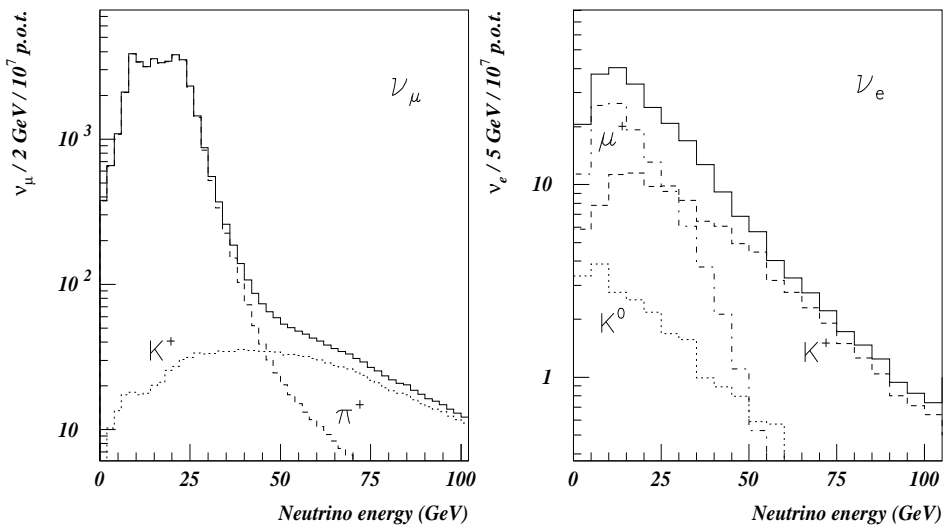,width=14.0cm,height=15cm}}
  \end{center}
  \vskip -7cm
  \caption{\sf Muon and electron neutrino flux spectra of the CNGS beam at the Gran Sasso
           Laboratories.}
  \label{cngs-beam}
 \end{figure*}

 \noindent The CNGS $\nu_\mu$  beam has been optimized for the $\nu_\mu \rightarrow \nu_\tau$
 appearance search.
 The beam-line design was accomplished
 on the basis of the previous experience with the WANF beam at CERN SPS \cite{WANF}.
 The expected muon neutrino flux at the Gran Sasso site will have
 an average energy of $17.4$ GeV and $\sim 0.6 \%$ $\nu_e$ contamination
 for $E_\nu < 40$ GeV (Fig. \ref{cngs-beam}). Due to the long-baseline  (L=732 Km) the contribution
 to neutrino beam from the $K^0$ and mesons produced in the reinteraction processes
 will be  strongly reduced with respect to the WANF \cite{WANF1}: the $\nu_e/\nu_{\mu}$ ratio
  is expected to be known within  $\sim 3 \%$ systematic uncertainty \cite{cngs_syst}.

 Current long-baseline experiments  with conventional neutrino beams  can look for
 $\nu_{\mu} \rightarrow \nu_e$ even if they are
 not optimized for $\theta_{13}$ studies.
  MINOS at NuMI is expected to reach a sensitivity of $\sin^2{2\thetaot}=0.08$
 \cite{Minos} integrating $14{\cdot}
 10^{20}$ protons on target (pot)
  in 5 years according to the FNAL proton plan evolution \cite{FermilabProtons}.
  MINOS main limitation
 is the poor electron identification efficiency of the detector.
 ICARUS \cite{ICARUS} and OPERA \cite{OPERA} can reach a 90\% C.L. combined sensitivity
  $\sin^2 2 \theta_{13}=0.030$ 
 ($\Delta m^2_{23} = 2.5 {\cdot} 10^{-3}$  eV$^2$, convoluted to CP and matter effects),
 a factor $\sim 5$ better than Chooz
 for five years exposure to the
 CNGS beam at nominal intensity for shared operation
  $4.5 {\cdot} 10^{19}$ pot/yr  \cite{Komatsu}.
 Depending on the $\delCP$ value and matter effects
 (\sigdm), these sensitivities can be reduced by a
 factor $\sim 1.5$ \cite{Migliozzi}.
 According to  the CERN  PS and SPS upgrade studies
 \cite{SPS_pot_increase},
 the CNGS beam intensity could be improved by a factor 1.5, allowing for
  more sensitive neutrino  oscillation searches for ICARUS and OPERA experiments.

 \noindent It is worth mentioning that
  the sensitivity on $\theta_{13}$ measurement of the current long-baseline experiments with
 conventional neutrino beams, like NUMI and CNGS, will be limited by the power of the proton source
 which determines the neutrino flux and the event statistics, by the not optimized $L/E_\nu$
 and by the presence of the $\nu_e$ intrinsic beam contamination and
 its related systematics.
 This is particular true for CNGS where the neutrino energy,
 optimized to overcome the kinematic
 threshold for $\tau$ production and to detect the $\tau$ decay products, is about
 ten times higher the optimal value for \thetaot searches.

 Another approach to search for non vanishing \thetaot is to look at \nubare\  disappearance
 using nuclear reactors as neutrino source. A follow-up of Chooz, Double Chooz \cite{DChooz}, has been
 proposed to start in 2008 with a two detectors setup, capable to push systematic errors down to 0.6 \%
 and to reach a sensitivity on $\sin^2{2\thetaot \simeq 0.024}$  (90\%
 C.L., $\dmtt=2.5{\cdot} 10^{-3})$ in a 3 years run.

\begin{figure}[!htp]
  \begin{center}
  \mbox{\epsfig{file=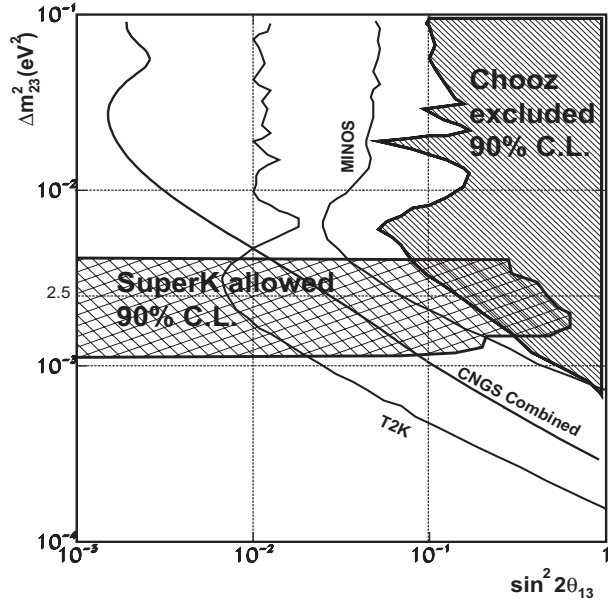, width=8.0cm, height=8cm}}
  \end{center}
  \vskip -0.5cm
  \caption{\sf Expected sensitivity on $\theta_{13}$ mixing angle
               (matter effects and  CP violation effects not included)
               for MINOS, ICARUS and OPERA combined at nominal CNGS
               and for the next T2K experiment,
               compared to the Chooz exclusion plot.}
  \label{exclusion}
 \end{figure}

 A summary of \thetaot sensitivities for the present generation of experiments
 is reported  Tab.~\ref{limiti1} and in Fig.~\ref{exclusion}.
 The same data are reported in Fig.~\ref{fig:th13vstime} as a function
 of the time, following the schedule reported in the experimental proposals.

 \begin{table}[ht]
 \caption{\sf The expected $90 \%$ C.L. sensitivity on $\theta_{13}$ measurements
         for the present and next generation
         long-baseline experiments with conventional $\nu_\mu$ beams for
         $\Delta m^2_{23} \sim 2.5 {\cdot} 10^{-3}$  eV$^2$ ($\delCP = 0$).
         The result of the reactor experiment  Chooz is also shown as comparison.}
 \begin{center}
 \begin{tabular}{l r  r r}
 \hline
 Experiment & fid. mass (Kt)& $\sin ^2 2 \theta_{13}$ & $\theta_{13}$ \\
 \hline
 MINOS               & 5.0         &  0.08   &$8.1^\circ$ \\
 ICARUS              & 2.4         &  0.04   &$5.8^\circ$\\
 OPERA               &  1.8        &  0.06   &$7.8^\circ$   \\
 Chooz               & 0.012       &  0.14   &$11^\circ$  \\
 \hline
 \end{tabular}
 \end{center}
 \label{limiti1}
 \end{table}

\section{The future experimental challenge: the sub-leading \numunue\  oscillations}
\label{section:numunue}

The parameters \thetaot, \delCP and \sigdm can be extracted by measuring sub-leading
\numunue\ oscillations. The
\nue\ disappearance experiments, like reactor experiments, can address
\thetaot searches but are not sensitive to \delCP and {{\rm sign}($\Delta m^2_{23})$}, while
\nuenutau\  transitions could provide similar information of \numunue\
transitions but they  are experimentally very challenging.

The \numunue\  transition probability can be parameterized as \cite{Richter}:
\begin{equation}
\begin{split}
      P&(\nu_\mu  \rightarrow \nu_e) =
 4 c^2_{13} { s^2_{13}} s^2_{23} \sin^2{\frac{\Delta m^2_{13}L}{4E_\nu}} \\
&
 +8c^2_{13}s_{12} s_{13} s_{23} (c_{12}c_{23} { cos\delCP}-s_{12}s_{13}s_{23})
   \cos{\frac{\Delta m^2_{23} L}{4E_\nu}}\sin{\frac{\Delta m^2_{13} L}{4E_\nu}}\sin{\frac{\Delta m^2_{12} L}{4E_\nu}}  \\
&
 -8c^2_{13} c_{12} c_{23} s_{12} s_{13} s_{23} {\sin{\delCP}}
   \sin{\frac{\Delta m^2_{23} L}{4E}}\sin{\frac{\Delta m^2_{13} L}{4E_\nu}}\sin{\frac{\Delta m^2_{12} L}{4E_\nu}} \\
&
 +4 {s^2_{12}} c^2_{13} \{c^2_{13} c^2_{23}+s^2_{12}s^2_{23}s^2_{13}-2c_{12}c_{23}s_{12}s_{23}s_{13} cos\delCP\}
 \sin{\frac{\Delta m^2_{12} L}{4E_\nu}}  \\
&
  -8c^2_{12}s^2_{13}s^2_{23}
   \cos{\frac{\Delta m^2_{23} L}{4E_\nu}}\sin{\frac{\Delta m^2_{13} L}{4E_\nu}}{\frac{aL}{4E_\nu}}(1-2s^2_{13}).
\end{split}
\label{eq:sub}
\end{equation}

\noindent The first line of this parameterization contains the term driven by \thetaot,
the second and third contain CP even and odd  terms respectively, and
the forth is driven by the solar parameters. The last line
parameterizes matter effects developed at the first order where
$a[{\rm eV}^2]=\pm 2\sqrt{2}G_Fn_eE_\nu=7.6{\cdot} 10^{-5}\rho[g/cm^3]E_\nu[{\rm GeV}]$.
The CP odd term and matter effects change sign by changing neutrinos with
antineutrinos.

\begin{figure*}
{\epsfig{file=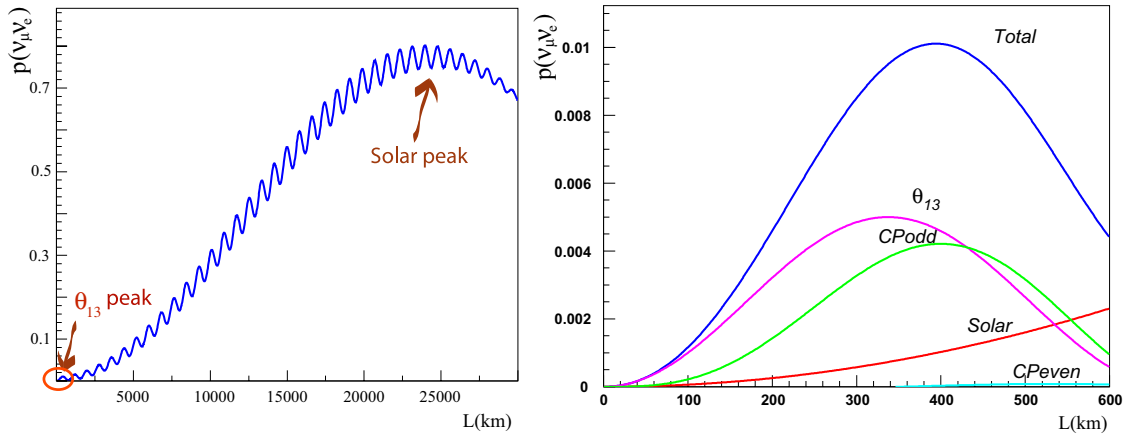,width=1.0\textwidth}}
    \caption{\sf Sketch of \pnumunue\  as function of the baseline computed for
             monochromatic neutrinos of 1 GeV in the solar baseline regime for \delCP=0 (left)
             and in the atmospheric baseline regime for $\delCP=-\pi/2$ (right),
             where the different terms of eq.~\ref{eq:sub} are displayed.
             The following oscillation parameters were used in both cases:
             $\sin^2{2\thetaot}=0.01$,
             $\sin^2{2\theta_{12}}=0.8$, $\Delta m^2_{23}=2.5{\cdot} 10^{-3}$ eV$^2$,
             $\Delta m^2_{12}= 7{\cdot} 10^{-5}$ eV$^2$.}
    \label{sub-leading}
\end{figure*}

\noindent The \numunue\  transitions are dominated by the solar term, anyway, at the distance
 defined by the $\Delta m^2_{23}$  parameter, they are driven by the \thetaot term which is
 proportional to $\sin^2{2\thetaot}$.
 Below $\sin^22\thetaot \simeq 10^{-3}$ the ``solar neutrino oscillation regime''
 will be  again the dominant transition mechanism, limiting
 further improvements of the experimental sensitivity
 to $\theta_{13}$.
 Moreover, $P(\nu_\mu  \rightarrow \nu_e)$ could be strongly
 influenced by the unknown value of \delCP and \sigdm.
  The contribution of the different terms of eq.~\ref{eq:sub}
  is shown in Fig.~\ref{sub-leading}
   as a function of the baseline for 1 GeV neutrinos 
   in the solar oscillation region (left) and in the
atmospheric oscillation region (right).

 \noindent  The measurement of $\theta_{13}$ represents the first mandatory ingredient
 for the investigation of the CP leptonic violation
 in the $\nu_\mu \rightarrow \nu_e$ transitions and for the mass
   hierarchy determination.
  The detection of the $\delta_{CP}$ phase will require a major experimental effort
  because of its intrinsic
 difficulty, relying on the need of disentangling several  contributions
 to $\nu_{\mu} \rightarrow \nu_e$ oscillation probability.

 \noindent Leptonic CP violation searches look for
 different behavior of neutrino and antineutrino appearance
 probabilities through the asymmetry

\begin{equation}
 A_{CP}=\frac{P(\numunue)-P(\nubarmunubare)}{ P(\numunue)+P(\nubarmunubare)}
\simeq \frac{\Delta m^2_{12} \, L}{4 \, E_\nu} {\cdot}
\frac{\sin 2 \theta_{12}}{\sin \theta_{13}} {\cdot} \delta_{CP}.
\end{equation}

\noindent The effect of the CP violation will be  proportional to $1/\sin\thetaot$
while \pnumunue\  is proportional to $\sin^22\thetaot$. For large values
of \thetaot, $A_{CP}$ will be small even if characterized by large number of
oscillated events: systematic
errors would be the main limiting experimental effect. For small values of
\thetaot event statistics and  background rates  would be the
ultimate limitation to the search.

\noindent Matter effects also produce differences between \pnumunue\  and \pnubarmunubare:
they depend from to the baseline and to the neutrino energy and could increase or decrease the
overall probabilities depending on
${{\rm sign}(\Delta m^2_{23})}$.
 At baselines of $\sim 100$ km these effects are negligible while
at $\sim 700$ km they can be up to $\sim 30\%$ of the probabilities in vacuum.

The richness of the \numunue\  transition is also its weakness: it will be very
difficult for pioneering experiments to extract all
the genuine parameters unambiguously. Correlations are present between \thetaot and
\delCP \cite{PilarNufact}. Moreover, in absence of information about
 the sign of $\Delta m^2_{23}$~\cite{Minakata,Barger:2001yr} and the
approximate $[\theta_{23}, \pi/2 - \theta_{23}]$ symmetry for the
atmospheric angle~\cite{Fogli}, additional clone solutions rise up. In general,
the measurement of $P(\nu_\mu \to \nu_e)$ and $P(\bar \nu_\mu \to \bar
\nu_e)$ will result in eight allowed regions of the parameter space,
the so-called eightfold-degeneracy~\cite{Barger:2001yr}.

 \vskip12pt

 \thetaot  searches look for experimental evidence
 of \nue\   appearance
 in excess to what expected from the solar terms.
 These measurements will be
 experimentally hard because the Chooz limit on the $\overline{\nu}_e$
 disappearance, $\theta_{13} < 11^\circ$ for $\Delta m^2_{23} \simeq 2.5 {\cdot} 10^{-3}$ eV$^2$,
 translates into a $\nu_{\mu} \rightarrow \nu_e$ appearance probability less than $10 \%$
 at the appearance maximum in a high energy muon neutrino beam.
 Furthermore, as already pointed out, the $\nu_{\mu} \rightarrow \nu_e$ experimental sensitivity
 with conventional $\nu_\mu$ beams is limited by an
 unavoidable $\nue$ beam contamination of about 1\%. The $\nu_{\mu}$ to $\nu_{\tau}$
  oscillations, with $E_\nu$ above the $\tau$ mass production threshold,
  generate background due to a significant number
  of $\nu_{\tau}$ charged current interactions
  where a large fraction of $\tau$'s decay into electrons.
   Finally, neutral pions in both neutral current or charged current interactions
  can fake an electron providing also a possible background for the $\nu_e$'s.

  \noindent Therefore the measurement of $\theta_{13}$ mixing angle and the
  investigation of the leptonic CP violation will require:

  \begin{itemize}

  \item [-] neutrino beams with high performance in terms of intensity, purity
            and low associated systematics. Event statistics, background rates
            and systematic errors will play a decisive role in detecting
            \nue\  appearance;

  \item [-] the use of detectors of unprecedent mass, granularity and
            resolution. Again event statistics is the main concern, while high
            detector performances are necessary to keep the event
            backgrounds (as $\pi^\circ$ from \numu\  neutral current interactions,
            mis-identified as \nue\  events) at  low as possible rate;

  \item [-] ancillary experiments to measure the meson production (for the
            neutrino beam knowledge), the  neutrino cross-sections, the
            particle identification capability. The optimization of
            proton driver characteristics and the best possible estimation of the
            systematic errors will require this kind of dedicated experiments.
            The Harp hadroproduction experiment at CERN PS \cite{Harp}
            took data for primary protons between 3 and  14.5 GeV in
            2001 and 2002 with different target materials.
            These data are expected to contribute to the proton driver optimization, the
            determination of the K2K and MiniBooNE neutrino beam fluxes and
            to the study of atmospheric neutrino interaction rates.
 \end{itemize}

 \noindent The intrinsic limitations of conventional neutrino beams are overcome
 if the neutrino parents can be fully selected, collimated and accelerated to a given energy.
 This  can be attempted within the muon or a beta decaying ion lifetimes.
 The neutrino beams from their decays would then be pure and perfectly predictable.
 The first approach brings to the Neutrino Factories \cite{Nufact},
 the second to the BetaBeams \cite{BetaBeam}.
 However, the technical difficulties associated
 with developing and building
 these novel conception neutrino beams  suggest for the middle
 term option to improve  the conventional beams by new high intensity
 proton machines,
 optimizing the beams for the $\nu_\mu \rightarrow \nu_e$ oscillation
 searches. They are called  SuperBeams \cite{Richter}.

 \noindent Different detection techniques of neutrino interactions based on
 water Cerenkov (WC), liquid Argon (LAr), nuclear emulsions  and calorimetry are available
 to build very massive detectors according to the intrinsic neutrino beam characteristics,
 energy and composition as discussed in the following section.

\section{Massive neutrino detectors}

Several experimental techniques have been recently exploited  for
high energy  neutrino detection and new set-up are under
consideration to cope with the challenges of precision neutrino
physics at accelerators. Compared with other high energy detectors, they
must offer unprecedented fiducial masses, instrumented with cheap and
reliable active detector technologies guarantying high
granularity, good energy resolution and excellent particle identification capability.
The most relevant technologies developed so far are discussed in the
following and summarized in Tab.~\ref{tech_tab}.

\subsection{Water Cerenkov Detector}
Besides the observation of solar and supernovae neutrinos, the discovery
of neutrino oscillations in the atmospheric sector represents another
great success of the Cerenkov technique. The target material and the
possibility to instrument only the vessel surface make these
detectors relatively cheap, so that huge fiducial masses are
conceivable.

Charged tracks above the Cerenkov threshold emit about 390 photons/cm
with a wavelength between 300 and 700 nm. Light attenuation in water,
as measured in Super-Kamiokande, is 98 m \cite{Fukuda:2002uc}.
Charged leptons are identified through the detection of Cerenkov
light, exploiting the features of the rings for particle
identification. A muon scatters very little crossing the detector.
Therefore, the associated ring has very sharp edges. Conversely, an
electron scatters (showers) much more, producing rings with ``fuzzy''
edges. 
 The total measured light gives
an estimate of the lepton energy, while the time measurement provided
by each photomultiplier determines the outgoing lepton direction
and the position of the neutrino interaction vertex. By
combining all this information it is possible to fully reconstruct the
energy, the direction and the flavor of the incoming neutrino. It is
worth noting that the procedure discussed above is suitable only for
quasi-elastic events ($\nu_l n \rightarrow l^- p$). Indeed, for non
quasi-elastic events there are other particles in the final state,
carrying a large energy fraction, that are either below the Cerenkov
threshold or neutrals, resulting in a poor
measurement of the total event energy.
Furthermore, the presence of more than one
particle above threshold produces more than one ring, spoiling the
particle identification capability of the detector.

In the SNO experiment a Cerenkov detector using heavy water as target
is employed for the detection or solar neutrinos.
Besides the features discussed above, the SNO
detector is also able to identify neutral current neutrino interactions
\cite{Boger:1999bb}  through the detection of the
neutron produced in the reaction $\nu_l d\rightarrow \nu_l p n$.

Water Cerenkov is a mature technology that demonstrated a valuable
cost effectiveness and excellent performances at low neutrino
energies. A detector with a fiducial mass as large as 20 times 
Super-Kamiokande \cite{UNO,T2K} could be built and would be an optimal
detector for neutrino beams with energies around or below 1 GeV. Furthermore,
such a device could represent
the ultimate tool for proton decay searches, and for atmospheric neutrinos
and supernovae neutrinos detection.


\subsection{Magnetized Iron Calorimeter}
\label{irondet}

Magnetized iron calorimeters are used since the seventies when the
HPWF experiment discovered neutrino induced charm-production
\cite{hpwf}. Calorimeters consist of magnetized iron slabs and
tracking detectors that act both as target and muon spectrometer. The
main feature of these detectors is their excellent muon reconstruction
(charge and momentum) and their high density (short interaction
length) that minimizes the background due to pion and kaon decays, and
to punch-through hadrons.

The MINOS Collaboration has built a magnetized iron calorimeter to
study neutrino oscillations at the atmospheric scale by using the NuMI
long-baseline beam~\cite{Minos}. The detector is composed of
2.54~cm-thick steel planes interleaved with planes of 1~cm-thick and
4.1~cm-wide scintillator strips. The iron is magnetized to an average
field of about 1.5~T. Simulations, as well as test beam results, show
that the energy resolution $\Delta E/E$ of this tracking calorimeter
is $55\%/\sqrt{E\mbox{({\rm GeV})}}$ and $23\%/\sqrt{E\mbox{({\rm GeV})}}$ for
hadronic and electromagnetic showers, respectively. This technology is
particularly suited for the measurement of $\nu_\mu$CC events, while
the electron identification is rather poor. Therefore, magnetized iron
calorimeters are planned to be used to study either $\nu_\mu$
appearance in a pure $\nu_e$ beam or $\nu_\mu$ disappearance in a well
known $\nu_\mu$ beam. It is worth noting that the presence of the
magnetic field is essential in order to tag the (anti-)neutrino in the
final state when such a detector is exploited in a non-pure beam,
i.e. conventional neutrino beams and Neutrino Factories (see
 Section~\ref{Nufact}). This technology has been proposed as a detector
for atmospheric neutrinos, Monolith \cite{Monolith}, and to study the
so-called ``golden channels''\cite{Cervera:2000kp}
$\nu_e\rightarrow\nu_\mu$ and $\bar{\nu}_e\rightarrow\bar{\nu}_\mu$ at
a Neutrino Factory~\cite{Cervera:2000vy}.

\subsection{Low Z Calorimeter}

Unlike the iron calorimeters discussed in Section~\ref{irondet}, low Z
calorimeters allow a good identification and energy measurement of
electrons produced in $\nu_e$ charged current interactions. In fact,
for this purpose one must sample showers more frequently than 
1.~X$_0$ and a magnetic field is not necessary. Another advantage of
a low Z calorimeter is that, for a given sampling in units of
radiation length, one can have up to a factor 3 more mass per readout
plane with respect to iron calorimeters. This detector can
discriminate between NC and $\nu_e$ induced charged current events by
looking at the longitudinal profile of the neutrino interaction, as
neutral current events are likely to be much more spread out in the
detector than $\nu_e$CC. Several active detectors (Resistive Plate
Chambers, streamer tubes, plastic and liquid scintillators) have been
considered and are currently under investigation. The liquid scintillator
 technique has been proposed by the NO$\nu$A experiment to search for
$\nu_{\mu}\rightarrow\nu_e$ oscillations in the NUMI Off-Axis
beam-line ~\cite {Nova}. The far detector will be composed solely of
liquid scintillator encased in a 15.7 m long cell titanium
dioxide-loaded PVC extrusions where each cell is 3.9 cm wide and 6 cm deep.

\begin{table*}
\begin{center}
 \renewcommand{\arraystretch}{1.3}
\begin{tabular}{||l|c|ccc|c|c||}
\hline
Detector Technology & Mass &
\multicolumn{3}{c|}{Event by event id} & Magnetic field &
 $\nu$ energy \\
        & (kton) & \nue & \numu & \nutau & & (GeV) \\
\hline
Water Cerenkov            & 50                    & OK & OK &   &          & 0.005 - 10 \\
Magnetic iron calorimeter & 5.4                   &    & OK &   & OK       & $>0.5$  \\
Low Z calorimeter         & 30 (project)          & OK & OK &   &          & 0.1-10 \\
Nuclear Emulsions (ECC)   & 1.8                   & OK & OK &OK & External & 1-100 \\
Liquid Argon TPC          & 0.6 $\rightarrow$ 3.0 & OK & OK &   & External & 0.001-100 \\
\hline
\end{tabular}
\end{center}
\caption{\sf Comparison of the main features of the experimental techniques
         discussed in the previous sections.}
\label{tech_tab}
\end{table*}

\subsection{Hybrid Emulsion Detector}

\begin{figure*}
\begin{center}
\includegraphics[width=4.in]{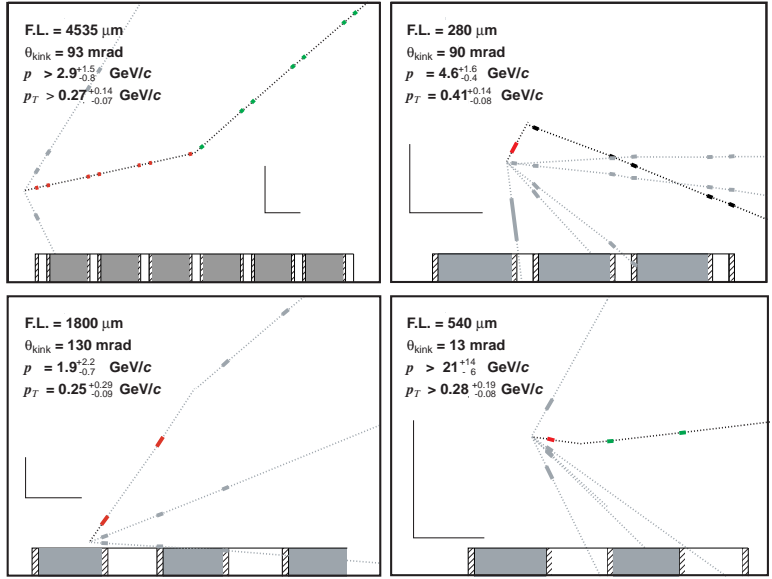}
\end{center}
\caption{\sf Schematic view of a $\tau$ decay candidate observed in the
ECC of the DONUT experiment. The neutrinos are incident from the left.
 The scale is given by perpendicular lines with the vertical line representing
 0.1~mm and the horizontal 1.0~mm. The target material is shown by the bar at the
 bottom of each part of the figure representing steel (dashed), emulsion 
(cross-hatched) and plastic (no fill, no shading). Top left: candidate
 $\tau\rightarrow e \nu_e\nu_\tau$; top right: candidate 
$\tau\rightarrow h \nu_\tau X$; bottom left: candidate $\tau\rightarrow h \nu_\tau X$;
 bottom right: candidate  $\tau\rightarrow e \nu_e\nu_\tau$.}
\label{donut}
\end{figure*}

The Emulsion Cloud Chamber (ECC) concept (see references quoted in
\cite{OPERA}), a modular structure made of a sandwich of
passive material plates interspersed with emulsion layers, combines the
high-precision tracking capabilities of nuclear emulsions and the
large mass achievable by employing metal plates as a target. By
assembling a large quantity of such modules, it is possible to
conceive and realize $\mathcal{O}(Kton)$ fine-grained vertex detector
optimized for the study of $\nu_\tau$ appearance. It has been adopted
by the OPERA Collaboration for a long-baseline
search of $\nu_{\mu}\rightarrow\nu_{\tau}$ oscillations at the CNGS
beam through the direct detection of the $\tau$'s produced
in $\nu_{\tau}$ charged current interactions. As an example of $\tau$
detection with ECC, we show in Fig.~\ref{donut} the $\nu_\tau$
events observed in the DONUT experiment~\cite{Kodama:2000mp}.

The basic element of the OPERA ECC is a ``cell'' made of a 1~mm thick
lead plate followed by a thin emulsion film which consists of
$44~\mu$m-thick emulsion layers on either side of a $200~\mu$m plastic
base. The number of grains hits in each emulsion layer ($15$-$20$)
ensures redundancy in the measurement of particle trajectories and
allows the measurement of their energy loss that, in the
non-relativistic regime, can help to distinguish between different
mass hypotheses.

Thanks to the dense ECC structure and to the high granularity provided
by the nuclear emulsions, the detector is also suited for
electron and $\gamma$ detection.  The energy resolution for
an electromagnetic shower is about 20\%. Nuclear emulsions
are able to measure the number of grains associated to each
track. This allows a two-track separation at  $\sim
1~\mu$m or even better. Therefore, it is possible to disentangle
single-electron tracks from electron pairs coming from $\gamma$
conversion in lead.
This outstanding position resolution can also be
used to measure the angle between different track segments with an accuracy of
about 1~mrad: this allows the use of Coulomb scattering to evaluate the
particle momentum with a resolution of about 20\%, and to 
reconstruct the kinematical  event variables.

A lead-emulsion detector has been also proposed to operate at a Neutrino
Factory to study the  ``silver channel''
$\nu_e\rightarrow\nu_{\tau}$ ~\cite{Silver,Autiero:2003fu} (see Section \ref{Nufact}).

\subsection{Liquid Argon Time Projection Chamber}

The technology of the Liquid Argon Time Projection Chamber (LAr TPC),
first proposed by C.~Rubbia on 1977~\cite{C.Rubbia}, was conceived as
a new tool for a completely uniform imaging with high accuracy of
very massive volumes, continuously sensitive and self-triggering.

\noindent The operating principle of this kind of detector is rather simple:
 any ionizing event (from a
particle decay or interaction) taking place in the active LAr
volume, which is maintained at a temperature $T \sim 89$~K, produces
ion-electron pairs. In the presence of a strong electric field ($\sim
0.5$~KV/cm), the ions and electrons drift.
The faster electrons are collected by  anode  wire planes with
different orientations located  near the end of the sensitive volume.
 The knowledge of the wire positions and the
drift time provides the three-dimensional image of the track, while
the charge collected on the wires provides precise information on the
deposited energy.

The detector developed by the ICARUS Collaboration~\cite{ICARUS},
consists of a large vessel of liquid Argon filled with three planes of wires
strung on the different orientations.
 The device allows tracking, $dE/dx$
measurements and a full-sampling electromagnetic and hadronic
calorimetry. Furthermore, the imaging provides excellent electron and
photon identification and electron/hadron separation. The energy
resolution $\Delta E/E$ is excellent for electromagnetic showers ($\sim
11\%/\sqrt{E\mbox{({\rm MeV})}}$, $E<50$ MeV  \cite{Icarus-muone},
$3\%/\sqrt{E\mbox{({\rm GeV})}}\oplus 1\%$ \cite{ICARUS})
 and also very good for contained hadronic
showers ($30\%/\sqrt{E\mbox{({\rm GeV})}}$). Furthermore, it is possible to
measure the momentum of muons with a resolution better than 20 \%
  by using multiple Coulomb scattering (3\%
for stopping muons where momentum is measured from range).

\begin{figure}
\begin{center}
\includegraphics[width=4.in]{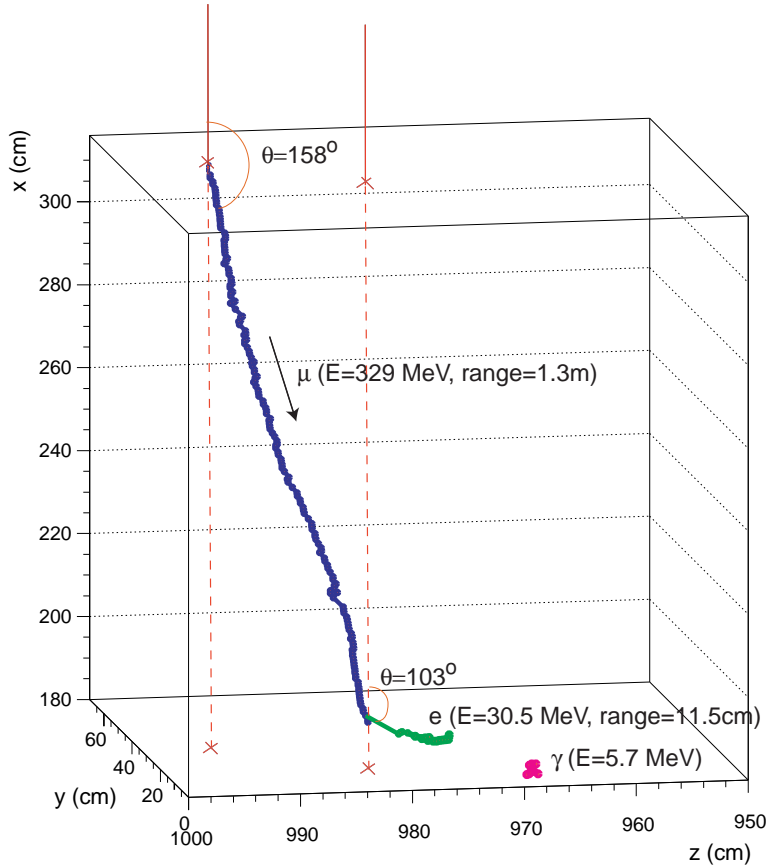}
\end{center}
\caption{\sf Stopping muon in the ICARUS 600~tons detector and decaying
into an electron.}
\label{icarus_eve}
\end{figure}

The most important milestone of this technique has been the successful
operation of the ICARUS 600~tons prototype \cite{600ton}, which has operated during
the summer of 2001 and it is now installed in the  Gran Sasso laboratories.
An event recorded with the 600~tons detector is shown
in Fig.~\ref{icarus_eve}.  A 3~kton detector is designed to operate with
the CNGS neutrino beam to search for $\nu_\mu\rightarrow\nu_\tau$ oscillations.
 Given its excellent electron identification
capabilities, it has been also proposed to operate on new other neutrino beams
to search for $\nu_\mu\rightarrow\nu_e$ appearance.

 A 100 kton LAr detector has been  proposed \cite{ARubbia} for a next future generation
 experiment which would deliver excellent physics output in rare event search and neutrino
 physics.
 However, new concepts and R\&D studies are required to extrapolate
 this technology further.
 The proposed detector would be composed by
 a single ``boiling'' cryogenic tanker, with external dimensions of
 40 m in height and 70 m in diameter.
 The charge imaging, scintillation and Cerenkov light readout allow a
 complete and redundant event reconstruction. The detector will be running in bi-phase mode,
 liquid plus gas, in order to allow for drift lengths as long as $\sim 20$ m along a
  drift electric field of $\sim 1 $ KV/cm.
 The drift electrons produced in the liquid phase are extracted
 from the liquid into the gas by a suitable electric field and then amplified
 near the anodes.

\section{New facilities for next generation of neutrino oscillation experiments}

  As pointed out above,  different options for neutrino beams of novel conception are presently
  under study for the next generation of the long-baseline neutrino oscillation
  experiments.   Different time scales can be envisaged according to the technical difficulties
  associated with the  developing and the building  these facilities.

  The first facility will probably be neutrino SuperBeams:
  conventional neutrino beams characterized
  by megawatt power proton drivers (Section \ref{SuperBeams}).
  Neutrino SuperBeams will require
  the development of high power proton Linacs or Rapid Cycling Synchrotrons, expected
  to happen in the next decade, and the development of proton targets able to survive to
  megawatt power proton beams, whose R\&D studies have already started \cite{targetry}.

  Neutrino SuperBeams can be seen as the injector stage of Neutrino Factories
  (Section \ref{Nufact}), where the other daughters of pion decays,
  the muons, are collected, cooled down,
  accelerated and stored in a decay ring where neutrinos are produced.
  The muon manipulation, acceleration and storage will require the development
  of novel machines in high energy physics accelerators with a consequent
  timescale of the order of about 20 years \cite{APS}.

  In BetaBeams (Section \ref{BetaBeam}) neutrinos are produced by the decay
  of radioactive heavy ions after proper acceleration.
   In principle all the machinery for
  ion production and acceleration has been already developed at CERN for the heavy ion
  physics program at Isolde and SPS. The required improvement by about 3 orders
  of magnitude of the presently available ion fluxes will require sub-megawatt, 1-2 GeV
  Linacs, new target developments for heavy ion production, ion
  collection and acceleration system including the CERN PS and SPS and a novel decay ring.
  Accounting for the technical challenges involved in the above facilities,
  the expected timescale of BetaBeams could be  intermediate between SuperBeams and
  Neutrino Factories. A design study of
  BetaBeam is just started within the European project Eurisol \cite{Eurisol} aimed at
  a very intense source of heavy ions.

\subsection{Near-term long-baseline experiments with SuperBeams}
\label{SuperBeams}

  According to the present experimental situation, conventional neutrino beams can be improved
  and optimized for the $\nu_{\mu} \rightarrow \nu_e$ searches.
  The design of a such new SuperBeam facility for a  very high intensity and low energy
  $\nu_\mu$ flux will demand:

 \begin{itemize}
   \item a new  higher power proton driver, exceeding the megawatt,
         to deliver more intense proton beams on target;
   \item a tunable $L/E_{\nu}$ in order to explore the $\Delta m^2_{23}$
         parameter region as indicated by the previous experiments with neutrino
         beams and atmospheric neutrinos;
   \item narrow band beams with $E_{\nu} \sim 1\div 2$ GeV;
   \item a lower intrinsic $\nu_e$ beam contamination which can be obtained
         suppressing the $K^+$ and $K^0$
         production by the primary proton beam in the target.
 \end{itemize}

 \noindent An interesting option for  the SuperBeams is the possibility to tilt the beam
 axis a few degrees with respect to the position of the far detector (Off-Axis beams)
 \cite{T2K, OffAxis}.
 According to the two body $\pi$-decay kinematics,   all the pions above a given
 momentum
 produce neutrinos of similar energy at a given angle $\theta \ne 0$
 with respect to the direction of parent pion (contrary to the $\theta=0$ case where
 the neutrino energy is proportional to the pion momentum).
 \noindent These neutrino beams have several advantages with respect to the
 corresponding on-axis ones:
 they are narrower, lower energy and with a smaller \nue\
 contamination (since \nue\  mainly come from three body decays)
 although the neutrino flux can be significantly smaller.

\begin{figure}[!htp]
  \begin{center}
  \mbox{\epsfig{file=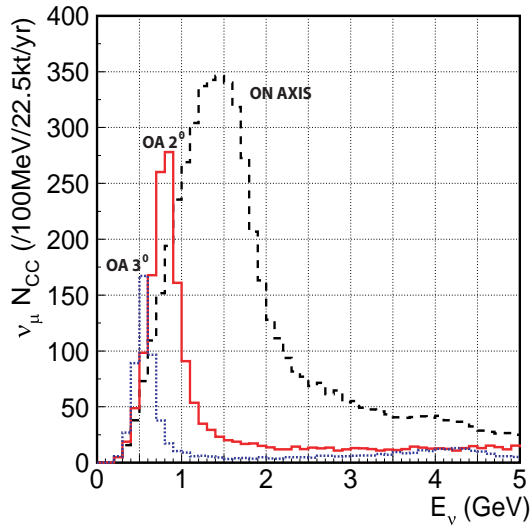, width=7cm, height=7cm}}
  \end{center}
  \vskip -0.5cm
  \caption{\sf T2K neutrino beam energy spectrum for different
               off-axis angle $\theta$.}
  \label{off-axisJHF}
 \end{figure}

 In the JHF project  Phase I (T2K experiment \cite{T2K}) a
 50 GeV/c proton beam of 0.75 MW from a PS will produce a very
 intense $\pi$ and $K$ beam tilted by $\theta=2^\circ$ with respect to the
 position of Super-Kamiokande detector at 295 Km of distance.
 The experiment is approved: data taking is scheduled to start in 2009 (10 \%
 of the project's pot intensity is expected the first year of run). The resulting
 $~ 700$ MeV $\nu_{\mu}$ beam (Fig. \ref{off-axisJHF}) with  $0.4 \%$ $\nu_e$
 contamination will allow to reach  a 90 $\%$ C.L. sensitivity $\sin^2 2 \theta_{13} \sim 0.006$
 in five years, assuming
 $\delta_{CP} = 0$, a factor 20 better than the current limit set by Chooz,
 see Fig.~\ref{exclusion1}.
 T2K will also measure \dmtt and $\sin^2{2\thetatt}$
 with  $\sim 2\%$ precision detecting \numu\    disappearance 
 and will perform a sensitive search for sterile neutrinos through the detection
 of neutral current event disappearance. 

\begin{figure*}[htb]
\centerline{\epsfig{file=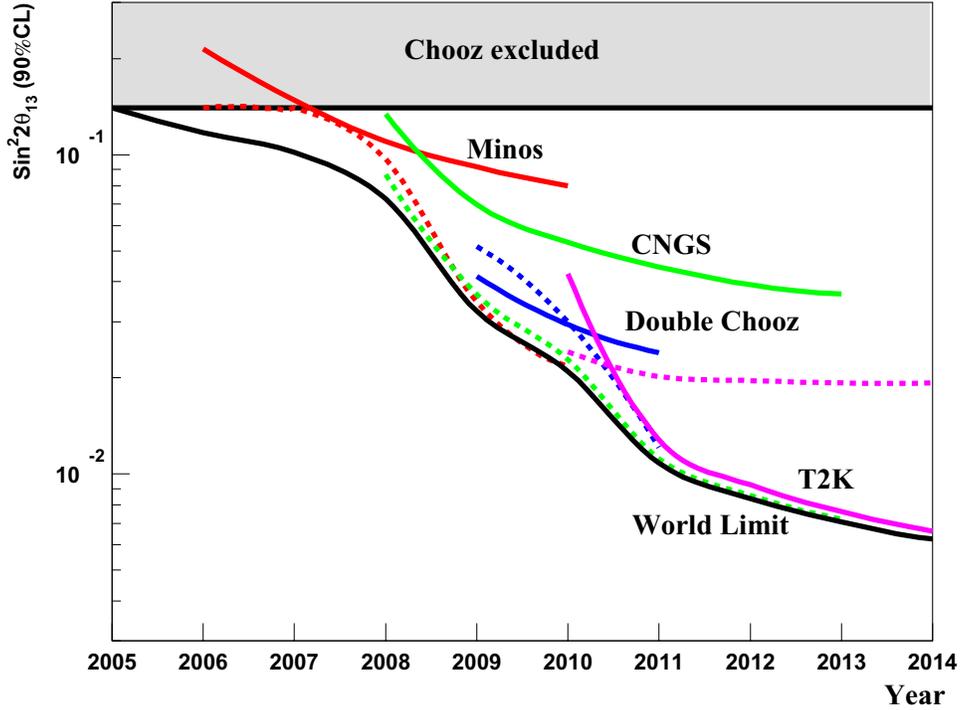,width=0.85\textwidth}}
    \caption{\sf Evolution of sensitivities on $\sin^2{2\thetaot}$
    as function of time. For each experiment are displayed the sensitivity as
    function of time (solid line) and the world sensitivity computed without
    the experiment (dashed line). The comparison of the two curves shows the discovery potential
    of the experiment  along its data taking. The world overall sensitivity along the
    time is also displayed. The comparison of the overall world sensitivity
    with the world sensitivity computed without a single experiment
    shows the impact of the results of the single experiment.
    Experiments are assumed to provide results
    after the first year of data taking.}
    \label{fig:th13vstime}
\end{figure*}

 \noindent The T2K sensitivity compared to the ones of the
 present generation experiments,
 is reported in Fig.~\ref{fig:th13vstime} as a function of the time.

 The foreseen machine upgrade to 4 MW (JHF-II), in conjunction with
 the construction of a very large (0.54 Mton fiducial volume) water Cerenkov
 detector (Hyper-Kamiokande) will allow to investigate the CP
 violation phase (J-Parc II). In a 5 years run with the \numu beam, the experiment could reach
 a 90\% C.L. $\thetaot$ sensitivity $\sin^2{2\thetaot}>6{\cdot} 10^{-4}$,
 while with 2 years of $\nu_{\mu}$
 and 6 years of $\overline{\nu}_{\mu}$ operations, it will discover a non vanishing \delCP at 
 a $3 \sigma$ level or better if $\sin |\delta_{CP}| > 20^\circ$
 and $\sin^2 2 \theta_{13} \sim 0.01$ \cite{Takashi} (see also Fig.\ref{fig:CP:delta}).

 The NO$\nu$A experiment with an upgraded NuMI Off-Axis neutrino beam
 \cite{Nova} ($E_{\nu} \sim 2 $
 GeV and a $\nu_e$ contamination less than $0.5 \%$) and with a baseline of
  810 Km (12 km Off-Axis), was recently proposed at FNAL with the aim to explore the
 $\nu_{\mu} \rightarrow \nu_e$ oscillations with a  sensitivity 10 times better
 than MINOS.
 If approved in 2006 the experiment could start data taking in 2011.
 The NuMI target will receive
 a 120 GeV/c proton flux with an expected intensity of $6.5 {\cdot} 10^{20}$ pot/year (
 $2 {\cdot} 10^7$ s/year are
 considered available to NuMI operations while the other beams are normalized to
 $10^7$ s/year).
 The experiment will use a near and a far detector, both  using liquid scintillator.
 In 5 years of data taking, with 30 Kt active mass far detector a sensitivity on
 $\sin^2 2 \theta_{13}$ slightly better
 than  T2K, as well as a precise measurement of  $|\Delta m_{23}^2|$ and 
 $\sin^2 2 \theta_{23}$, can be achieved.
 NO$\nu$A can also allow to solve the mass hierarchy problem for a limited
 range of the \delCP and \sigdm parameters \cite{Nova}.

 \noindent As a second phase, the new proton driver of
 8 GeV/c and 2 MW, could increase the  NuMI beam
  intensity to $17.2 \div 25.2 {\cdot} 10^{20}$ pot/year,
 allowing to improve the experimental sensitivity by a factor two 
 and to initiate the experimental search for the CP violation.

 A longer term experiment has been proposed at BNL for a different
 long-baseline neutrino beam \cite{Diwan}. In this proposal the AGS 28 GeV PS
 should be upgraded to 1 MW and a neutrino beam with $\langle E_\nu \rangle
\simeq 1.5$ GeV should be fired into a megaton water Cerenkov detector
 at a baseline of 2540 km. The detector would
 be at the second oscillation maximum and the comparison of \numu
 disappearance and \nue\  appearance at the first and second oscillation
 maximum could allow a better control of degeneracies.
 However, it should be noted that background rates and signal efficiency
 of a water Cerenkov detector in this energy range are not optimal and
 not constant between the first and the second maximum.
 In 5 years run this experiment could reach a 90 \% C.L.
 sensitivity $\sin^2{2\thetaot}\simeq 0.003$ ($\delta_{CP} =0$).

 The performances of future neutrino oscillation experiments are summarized in Tab.~\ref{tab:confr}
 (see also Refs.~\cite{SBall}).

\subsubsection{European Superbeam projects}

 Many ideas and approaches have been developed for neutrino
 long-baseline experiments in Europe after the CNGS $\nu_{\tau}$ appearance
 phase.  These projects are aiming to improve and develop existing infrastructures
 and detectors or considering new neutrino beams and detectors.

 \vskip 12pt

\begin{figure}[!htp]
  \begin{center}
  \mbox{\epsfig{file=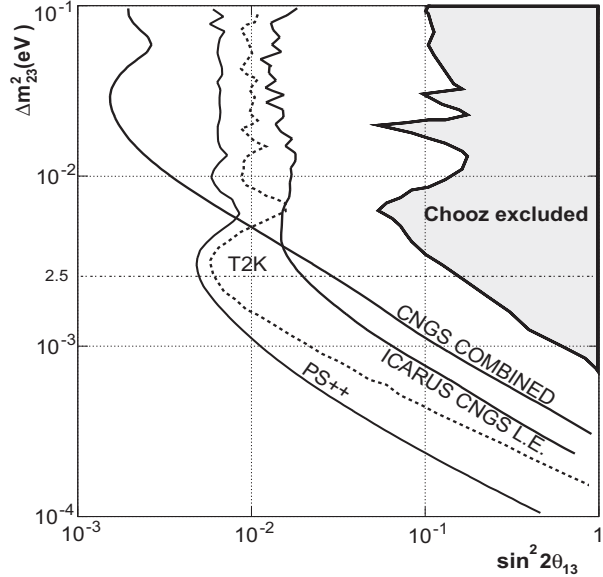,width=9.0cm, height=8.5cm}}
  \end{center}
  \vskip -0.5cm
  \caption{\sf Expected sensitivity on $\theta_{13}$ mixing angle
               (CP violation and matter effects not included in the
               calculation) for a 20 GeV/c high intensity PS proton
               beam from CERN to Gran Sasso (PS$++$) and for the ICARUS 2.4 Kton at the
               CNGS-L.E. compared to T2K experiment.}
  \label{exclusion1}
 \end{figure}

 The possibility to improve the CERN to Gran Sasso neutrino beam performances
 for $\theta_{13}$ searches even with the present proton beam, $E_p =  400$ GeV and
 $4.5 {\cdot} 10^{19}$ pot/year was  investigated (CNGS-L.E.) \cite{CERN-LOW}. 
 The low energy neutrino
 flux can be increased by a factor 5 with respect to the current CNGS beam by
 an appropriate optimization of the target (a compact 1 m carbon rod) and
 of the focusing system.
 \noindent The decay tunnel will be reduced to 350 m
 allowing for a near detector useful for determining beam composition.
 This intense low energy neutrino flux, $E_{\nu_{\mu}} \simeq 1.8$ GeV,
 will produce  4.5 $\nu_{\mu}$-CC$/10^{19}$
 pot$/$Kt event rate with  $0.9 \%$ $\nu_e/\nu_{\mu}$-CC event contamination.
 With this low energy CNGS-L.E. neutrino beam, the sensitivity
 to  $\sin^2 2 \theta_{13}$ can be increased by a factor 7 with respect to  Chooz,
  $\sin^2 2 \thetaot < 0.02$
 (not accounting for CP violation and matter effects) in 5 years with a 2.4 Kt fiducial
 mass liquid Argon detector by assuming $\Delta m^2_{23} = 2.5 {\cdot} 10^{-3}$ eV$^2$,
 Fig. \ref{exclusion1}.
 However, the preparation of a such low energy neutrino facility 
 wouldn't be compatible with the CNGS beam for \nutau programme and
 will require a suitable interval of time after the CNGS $\tau$-phase running for
 the "radioactive cooling" of the target and of decay tunnel in order to 
 change the target geometry, shorten and enlarge the decay tunnel,
 in a safe environment.
 Alternatively a new beam line should be build.

 A second study considered again a low energy neutrino beam (1.5 GeV mean
 energy) fired into a detector made of 44,000 phototubes deployed 
 in the Golfo di Taranto at 1000 m depth,   1200 km from CERN (CNGT), 2$^\circ$ 
 degrees off-axis, equipping 2 Mton of water \cite{CNGT}.
 In this case the detector would be placed at the second oscillation
 maximum and if movable it could take data both at the minimum and at the
 maximum of oscillation probability. Sensitivity would be marginally worse than
 T2K in a 5 years data taking \cite{CNGT}.

\vskip 0.5cm

 \noindent A proton driver was also recently studied to optimize the search for  the
 $\nu_{\mu}  \rightarrow \nu_e$ oscillations
 with a new generation of low energy and high intensity
 SuperBeam \cite{CarloRubbia}.
 In term of proton economics, the optimum beam energy turns out to be around 20 GeV,
 well matched to a 732 Km of baseline (i.e. CERN - Gran Sasso: average neutrino
 energy $E_\nu \sim 1.6 $ GeV). Roughly
 a $\sin^2 2 \theta_{13}\simeq 0.005$ sensitivity for
 $\Delta m^2_{23} \sim 2.5 {\cdot} 10^{-3}$ eV$^2$,
 assuming $\delCP=0$ and no
 matter effects,
 can be obtained for $2 {\cdot} 10^{22}$ pot/year
 (about two orders of magnitude higher than the intensity deliverable
 by the current CERN-PS) and 5 years exposure of ICARUS.
 The performance of this facility, indicated as PS$++$, has been computed for a power
 source corresponding to 6.5  MW accounting for a useful beam time operation
 of $10^7$ s per year and a 2.35 kton detector: anyway the same sensitivity can
 be reached with  4 MW power if a LNGS hall is fully occupied by ICARUS (about 4 kton).

\begin{figure*}[!htp]
  \epsfig{file=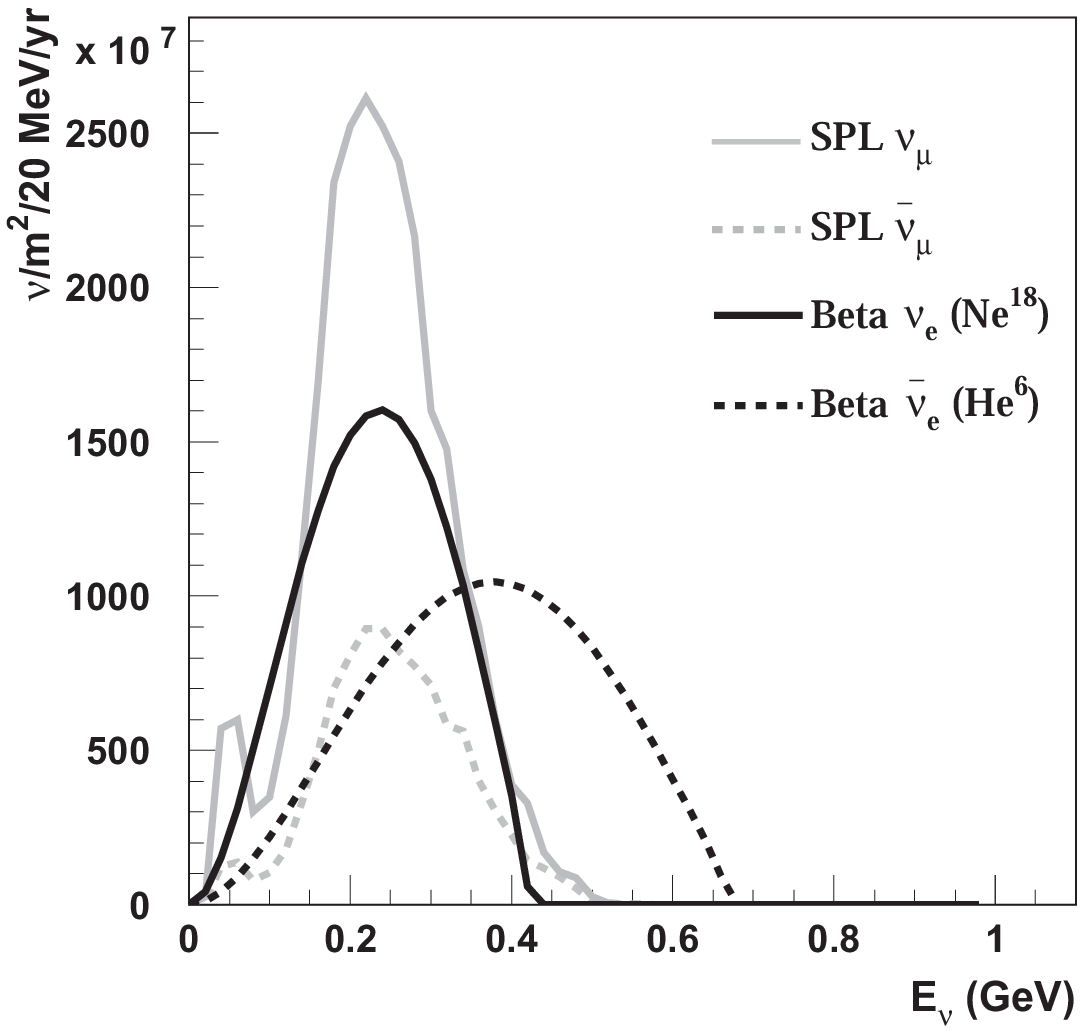,width=0.45\textwidth}
  \epsfig{file=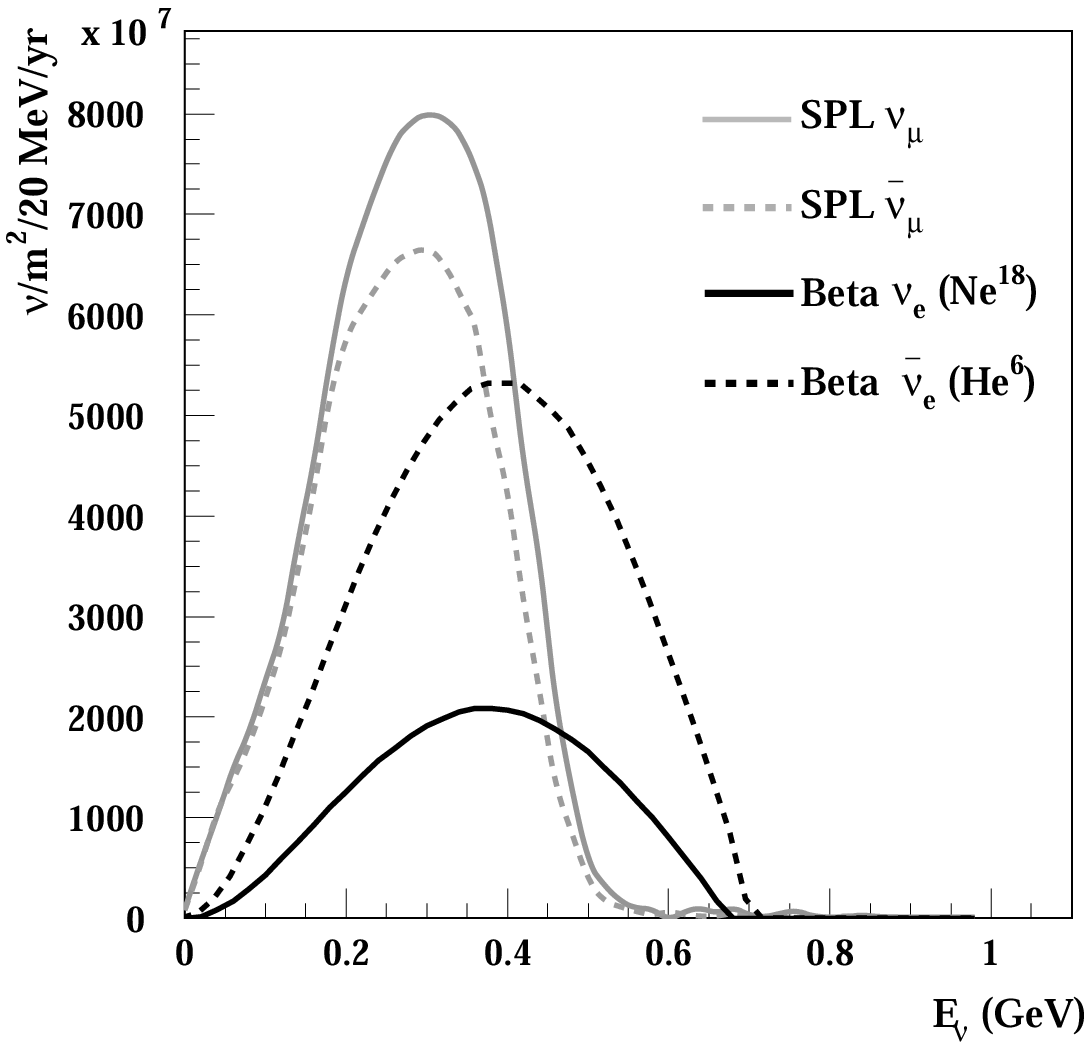,width=0.45\textwidth}
  \vskip -0.5cm
  \caption{\sf Left: neutrino flux of $\beta$-Beam ($\gamma_{\He}=60$,
  $\gamma_{\Ne}=100$, shared mode) and CERN-SPL SuperBeam,
           2.2 GeV, at 130 Km of distance.
           Right: the same for $\gamma_{\He}=100$,   $\gamma_{\Ne}=100$,
           (non shared mode, that is just one ion circulating in the decay
           ring)
           and a 3.5 GeV SPL SuperBeam.}
  \label{fig:fluxes}
 \end{figure*}

 In  the CERN-SPL SuperBeam project  \cite{SPL-Design,SPL-Physics,nufact1}
 the planned 4MW SPL (Superconducting Proton Linac)  would deliver a 2.2  GeV/c
 proton beam,  on a Hg target to generate
 an intense $\pi^+$ ($\pi^-$) beam focused by a suitable
 magnetic horn in a short decay tunnel. As a result   an intense
 $\nu_{\mu}$ beam, will be produced
 mainly via the $\pi$-decay,  $\pi^+ \rightarrow \nu_{\mu} \; \mu^+$ providing a
 flux $\phi \sim 3.6 {\cdot} 10^{11} \nu_{\mu}$/year/m$^2$  at 130 Km
 of distance, and an average energy of 0.27 GeV (Fig. \ref{fig:fluxes}).
 The $\nu_e$ contamination from $K$ will be suppressed by threshold effects
 and the resulting $\nu_e/\nu_{\mu}$ ratio ($ \sim 0.4 \%$) 
  will be known within  $2\%$ error.
 The use of a near and far detector (the latter at $L = 130$ Km of distance
 in the Frejus area)
 will allow for both $\nu_{\mu}$-disappearance and
 $\nu_{\mu} \rightarrow \nu_e$ appearance studies.
 The physics potential of the 2.2 GeV SPL SuperBeam (SPL-SB)
 with a water Cerenkov far detector fiducial mass of 440 Kt \cite{UNO}  has been extensively
 studied \cite{SPL-Physics}.
   The experimental sensitivity is displayed, due to the strong $\thetaot-\delCP$ correlation,
 in the (\thetaot-$\delCP$) plane having fixed $\dmtt=2.5{\cdot}10^{-3}$ eV$^2$ 
  (Fig.~\ref{fig:th13}).
 The 90 \% C.L. \thetaot sensitivity ($\delCP=0$) is  
$\sin^2{2\thetaot} = 0.002$ (5 years \numu beam), see Tab.~\ref{tab:confr}.
  The corresponding $3 \, \sigma$ CP
 violation discovery potential (2 years with
 $\nu_{\mu}$ beam and 8 years with the reversed polarity
 $\overline{\nu}_{\mu}$ beam) is shown in Fig.~\ref{fig:CP:delta}.

 New developments show that the potential of the SPL-SB potential could be
 improved by rising the SPL energy to 3.5 GeV \cite{Cazes},
 to produce   more copious secondary mesons
 and to focus them more efficiently. This seems feasible if
 status of the art RF cavities would be used in place of the old foreseen LEP cavities
 \cite{Garoby-SPL}.
 In this upgraded configuration neutrino flux could be increased by a factor 3 with
 with respect to the 2.2 GeV configuration, reaching
 a sensitivity to $\sin^2{2 \thetaot}$ 8 times better than T2K and allowing
 to discovery CP violation (at 3 $\sigma$ level) if
 $\delCP \geq 25^\circ$  and
 $\theta_{13} \geq 1.4^\circ$ \cite{MMNufact04}. The expected
 performances are shown in Fig.~\ref{fig:th13} and Fig.~\ref{fig:CP:delta}.

\subsection{BetaBeams}
\label{BetaBeam}

BetaBeams ($\beta$B)  have been introduced by
P. Zucchelli in 2001 \cite{BetaBeam}.
The idea is to generate pure, well collimated and intense
\nue\  (\nubare) beams by producing, collecting, accelerating radioactive ions
and storing them in a decay ring in 10 ns long bunches, to suppress
the atmospheric neutrino backgrounds.
The resulting $\beta B$ would be virtually background free and fluxes
could be easily computed by the properties of the beta decay of the parent
ion and by its Lorentz boost factor $\gamma$. The best ion candidates so far
 are  $^{18}Ne\;$  and $^6He\;$ for \nue\ and \nubare\  respectively.
\begin{figure*}[ht]
 \centerline{\epsfig{file=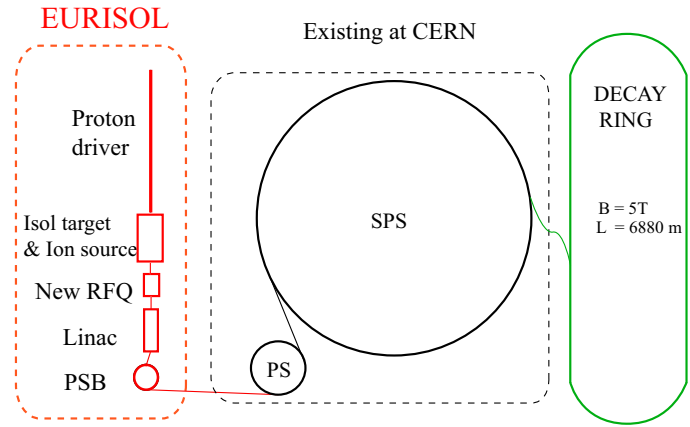,width=0.60\textwidth}  }
\caption{\sf A schematic layout of the BetaBeam complex. At left, the low energy part is
largely similar to the EURISOL project \cite{Eurisol}.
 The central part (PS and SPS) uses
existing facilities. At right, the decay ring has to be built.}
\label{fig:sketch}
\end{figure*}
Summarizing, the main features of a neutrino beam based on the
BetaBeams concept are:

\begin{itemize}
\item the beam energy depends on the $\gamma$ factor. The ion
accelerator can be tuned to optimize the sensitivity of the
experiment;
\item the neutrino beam contains a single flavor with an energy
spectrum and intensity known a priori. Therefore, unlike conventional
neutrino beams, close detectors are not necessary to normalize the fluxes;
\item neutrino and anti-neutrino beams can be produced with a
comparable flux;
\item ifferently from SuperBeams, BetaBeams experiments search for $\nu_e
\rightarrow \nu_\mu$ transitions, requiring a detector capable to
identify muons from electrons.
Moreover, since
the beam does not contain $\nu_\mu$ or $\bar{ \nu}_\mu$ in the initial
state, magnetized detectors are not needed. This is in contrast with
the neutrino factories (see below) where the determination of the muon
sign is mandatory.
\end{itemize}

\begin{figure*}[thb]
    \centerline{\epsfig{file=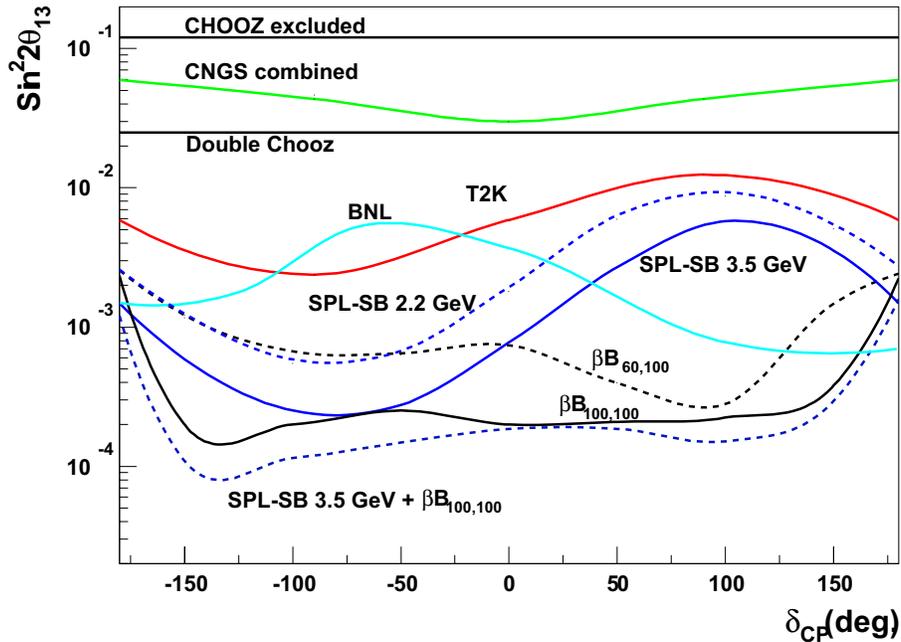,width=0.80\textwidth}  }
    \vspace{-0.5cm}
    \caption{\sf \thetaot \  90\%
             C.L. sensitivity as function of $\delCP$ for
             $\dmtt=2.5{\cdot}10^{-3}eV^2$, $\sigdm=1$, 2\%
             systematic errors.CNGS and T2K curves are taken from \cite{Migliozzi},
             BNL from \cite{Diwan}, Double Chooz from \cite{DChooz}.
             SPL-SB  sensitivities have been computed for a
             5 years \numu run, $\beta$B and $\beta$B$_{100,100}$
             for a 5 years \nue + \nubare\  run. }
  \label{fig:th13}
\end{figure*}

\begin{figure*}[thb]
    \centerline{\epsfig{file=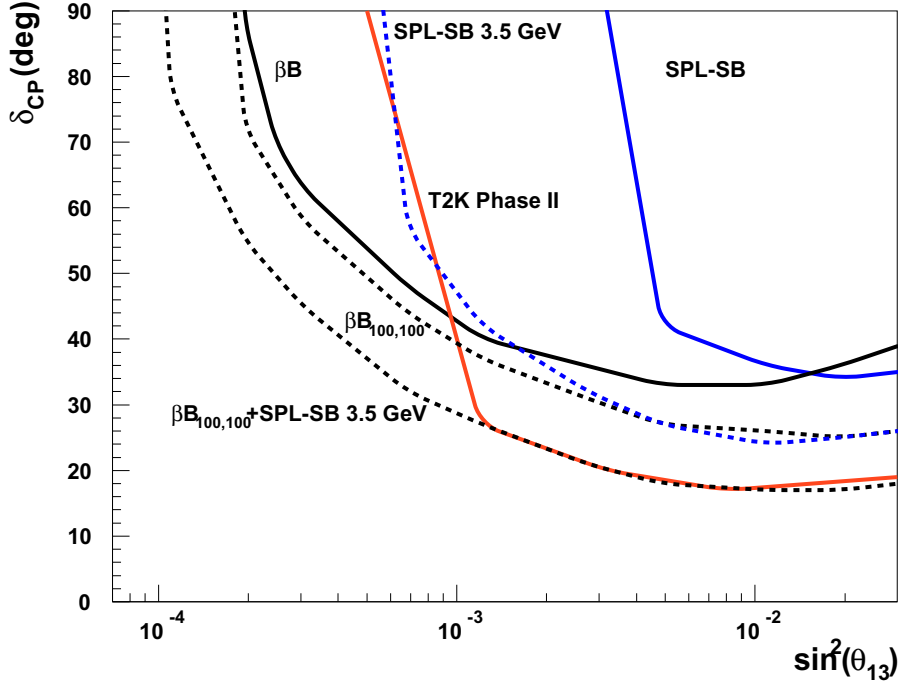,width=0.80\textwidth}  }
    \vspace{-0.5cm}
    \caption{\sf
      $\delCP$ discovery potential at $3 \sigma$ (see text) computed
      for 10 years running time.
      The SPL-SB 2.2 and 3.5 GeV, BetaBeam with $\gamma=60,100$ and
     $\gamma=100,100$ J-Parc II \cite{Takashi} and
      SPL-SB combined with BetaBeam are shown.
      All the curves are computed with a 2\% systematic error, 10
      years of data taking.}
  \label{fig:CP:delta}
\end{figure*}

\noindent A baseline study for  a Beta Beam complex (Fig.~\ref{fig:sketch})
has been carried out at CERN \cite{Lindroos}.
The SPS could accelerate \He\ ions at a maximum $\gamma$ value of
$\gamma_{\He}=150$ and \Ne\  ions up to $\gamma_{\Ne}=250$.
In this scenario the two ions circulate in the decay ring
at the same time. This is a feasible option, provided that their $\gamma$
are in the ratio $\gamma_{\He}/\gamma_{\Ne}=3/5$.
The reference $\beta$B fluxes  are $2.9 {\cdot} 10^{18}$ \He\ useful
decays/year and $1.1{\cdot}10^{18}$ \Ne\  decays/year if the
two ions are run at the same time in the complex.
The resulting neutrino fluxes are displayed in Fig.~\ref{fig:fluxes}.
The water \v{C}erenkov could be a suitable technology for a large detector.
The physics potential has been computed in \cite{beta} for $\gamma_{\He}=60$,
$\gamma_{\Ne}=100$ and
with a 440 kton detector at 130 km, they  are displayed in
Fig.~\ref{fig:th13} and Fig.~\ref{fig:CP:delta}.
Sensitivities taking into account all the parameter degeneracies and ambiguities 
 scenario have been computed in \cite{Donini:2004hu}.

Novel developments, suggesting the possibilities
of running the two ions separately at their optimal $\gamma$ \cite{MatsPrivate}, have
recently triggered a new optimal scheme for the BetaBeam.
In this scheme both ions are accelerated at $\gamma=100$.
The expected performances
are displayed ($\beta$B$_{100,100}$) in Figs.~\ref{fig:th13} and~\ref{fig:CP:delta}.
A sensitivity to $\sin^2{2 \thetaot}$ 30 times better than T2K could be reached and
lepton CP violation could be discovered at 3 $\sigma$ if $\delCP \geq 25^\circ$  and
$\theta_{13} \geq 1.0^\circ$ \cite{MMNufact04,latestJJ}.

BetaBeams require a proton driver in the energy range of 1-2 GeV, 0.5 MWatt power.
The SPL can be used as injector, at most 10\% of its protons would be consumed.
This allows a simultaneous $\beta$B and SPL-SB run, the two neutrino
beams having similar neutrino energies
(see also Fig.~\ref{fig:fluxes}). The same detector could then be
exposed to $2{\times} 2$ beams (\numu  and \nubarmu\  ${\times}$
\nue\  and \nubare) having access to CP, T and CPT violation searches in the same run.
With this combination of neutrino beams a sensitivity to $\sin^2{2 \thetaot}$
35 times better than T2K could be reached exploiting a CP violation discovery potential
at 3 $\sigma$ if $\delCP \geq 18^\circ$  and $\theta_{13} \geq 0.55^\circ$ \cite{MMNufact04}
(Figs.~\ref{fig:th13} and~\ref{fig:CP:delta}).

%
%

BetaBeam capabilities for ions accelerated at higher energies than
those allowed by SPS have been computed in
\cite{latestJJ,HighEnergy,HighEnergy2}. All these studies assume that
the same ion fluxes of the baseline scenario can be maintained.
However, this is not the case if the number of stored bunches is kept
constant in the storage ring. On the other hand, by increasing
$\gamma$ (i.e. the neutrino energy) the atmospheric neutrinos background
constraint on the total bunch length \cite{BetaBeam} tends to vanish.
Studies are in progress at CERN in order
to define realistic neutrino fluxes as a function of
$\gamma$\ \cite{MatsPrivate}. It is worth noting that if a high intensity Beta
Beam with $\gamma\sim 300\div500$ (requiring a higher energy
accelerator than SPS, like the Super-SPS\cite{SuperSPS}) can be built, a 40 kton
iron calorimeter located at the Gran Sasso Laboratory will have the
possibility to discover a non vanishing $\delta_{CP}$
if $\delCP>20^\circ$ for
$\theta_{13}\ge2^\circ$ (99\% C.L.) and measure the sign of $\Delta m^2_{23}$\cite{HighEnergy}.
%

\begin{table*}[htb]
\caption{\sf Summary table of different LBL options. J-Parc II $\sin^2{2\theta_{13}}$
 sensitivity is
extrapolated from T2K phase I. All the experiment are normalized to 5 years data taking considering
a neutrino beam time operation of $10^7$ s per year.
Numbers quoted for No$\nu$A refer to the standard and the proton driver options (see text).
SPL numbers are for the 2.2 GeV option (the 3.5 GeV performances are in parentheses).
The $\beta$B columns is computed for the $\gamma=60, \,100$ option (the $\gamma=100,\,100$ performances
are in parentheses), the $\nu${CC} line
indicates the sum of $\nue^{CC}$ and $\nubare^{CC}$ rates.
 The $\pi^\circ/\nue$ line indicates
the fraction of the neutral current background normalized to intrinsic  \nue\  background.  
Once fixed $L/E_\nu$ to well match the
 $\Delta m_{23}^2$ value, the figure of merit of the neutrino beam is determined by the
 $\nu_\mu$-CC/Kt/year event rate and also by the $\nu_e/\nu_\mu$ natural beam contamination.}
\label{tab:confr}
\vspace*{0.2cm}
 \renewcommand{\arraystretch}{1.3}
 \begin{tabular}{lc|ccccccc}
\hline
                                &       &  T2K    & J-Parc II  & NO$\nu$A &   BNL &  PS$++$  &   SPL (3.5) &$\beta$B ($\beta$B$_{100,100}$)\\
\hline
 p-driver                       & (MW)  & 0.75    &    4      &  0.8 (2)  &    1    &    4     &  4         & 0.4       \\
 p beam energy                  & (GeV) &  50     &    50     &  120      &   28    &    20    & 2.2 (3.5)  & 1-2.2     \\
 $\langle E({\nu})\rangle$      & (GeV) &  0.7    &    0.7    &  2        &   1.5   &    1.6   &0.27 (0.29) & 0.3  (0.4)\\
 $L$                            & (Km)  & 295     &   295     &  810      & 2540    &   732    &   130      & 130       \\
 Off-Axis                       &       &$2^\circ$&$2^\circ$  &0.8$^\circ$&  -      &    -     &    -       &     -     \\
 $\nu$ CC (no osc.)& $(Kt^{-1} yr^{-1}$)
                                        &  100    &   500     & 80 (200)  & 11      &   450    & 37 (122)  &  38  (56)   \\
 $\nu$ contamination            & (\%)  &  0.4    &   0.4     & 0.5       & 0.5     & 1.2      & 0.4 (0.7) &     0       \\
 \hline
Detect. Fid. Mass               & (Kt)&   22.5    &   540     &  30       &  440   &   3.8     &  440      &    440      \\
Material                        &       & H$_2$O  & H$_2$O    & LScint    & H$_2$O &  LAr      & H$_2$O    & H$_2$O      \\
Signal efficiency               & (\%)  &  40     &   40      &  24       &   25   &  100      &  70       &  60  (70)   \\
$\pi^\circ /\nu_e$ ($\pi/\numu$)& (\%)  &  80     &   80      &  60       &  100   &   0       &  30       &    -        \\
$\sin^2{2\theta_{13}}{\cdot} 10^4$
                                &(90\% C.L.)& 60  &    6      &  38 (24)  &  30    & 50        & 18  (8)   &   7  (2)    \\
\hline
\end{tabular}
\end{table*}

\subsection{Neutrino Factories}
\label{Nufact}

  The neutrino production by muon decay from a pure muon beam  has been considered
since 1998 \cite{Nufact}:
 this is indeed a perfectly well known weak process
 and the $\mu$ beam can be well measured in momentum and intensity.

 In the CERN present layout for a Neutrino Factory ({\large $\nu$F})
 \cite{nufact2} a
 4 MW proton beam  is accelerated up to 2.2 GeV/c by the Super Conducting Proton Linac
 (SPL) to produce  low energy $\pi$'s in a liquid mercury target,
 which are collected by a solenoid.
 Muons produced from the $\pi$-decay are then cooled and phase rotated before
 acceleration through a recirculating Linac system up to 50 GeV/c.
 These muons of well defined charge and momentum are
 injected in the $\mu$ accumulator where they will circulate until they decay,
 delivering along the two main straight sections two pure $\nu$ beams
 whose intensity is expected more than 100 times the one in
 conventional beams. Both muon signs can be selected. The decay
 $\mu^+ \rightarrow e^+ \nu_e \overline \nu_\mu$
 $ (\mu^- \rightarrow e^- \overline \nu_e \nu_\mu)$ produces
 a pure well collimated neutrino beam with equal numbers of
 $\overline \nu_\mu,\,\nu_e$ $\; (\nu_\mu,\, \overline{\nu}_e)$ and
 their energy allows to extend the baseline to  several thousand kilometers of
 distance.

\begin{figure}[!htp]
  \begin{center}
  \mbox{\epsfig{file=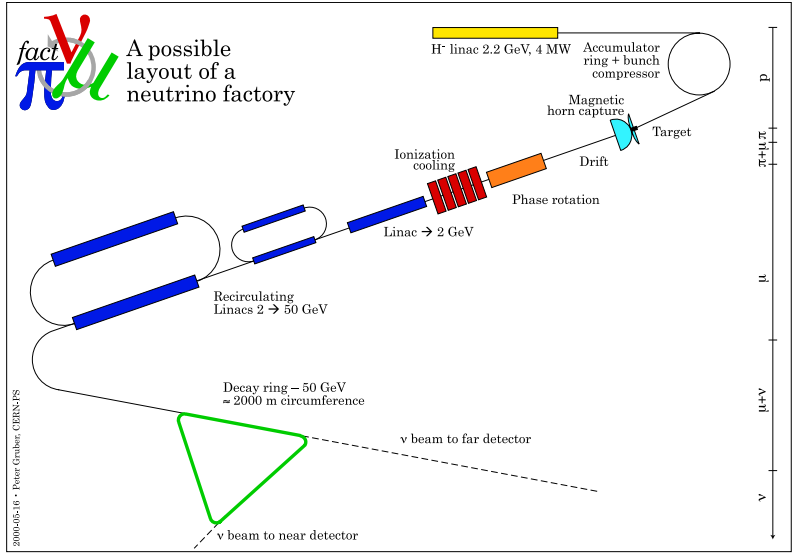,width=0.7\textwidth}}
  \end{center}
  \vskip -0.3cm
  \caption{\sf Expected layout for a neutrino factory at CERN.}
  \label{fig:nufact}
 \end{figure}

 \noindent The optimal beam energy at the {\large $\nu$F} will be as large
 as possible accounting for the difficulties and the technical challenge for the 
 construction of
 such a muon accelerator complex. $E_{\mu} = 50$ GeV ($E_{\nu} \sim$ 34 GeV) represents
 a limit value for  such a new machine.
 In fact the neutrino flux $\phi_{\nu}$ grows like $E_{\nu}^2$ (in the
 conventional neutrino beams $\phi_{\nu}$ is proportional to $E_{\nu}$);
 the number of charged current neutrino events from the
 oscillations ($N_{osc}$), measured by a detector at a distance L, will
 be proportional to $E_{\nu}$:

     \begin{equation}
              N_{osc} \propto \phi_{\nu} {\cdot} \sigma_{\nu} {\cdot} P_{osc}
              \propto \frac{{E_\nu}^3}{L^2} {\cdot} \sin^2{\frac{L}{E_{\nu}}}
              \simeq E_{\nu}
     \end{equation}

 \noindent where $\sigma_{\nu}\propto E_\nu$ is the corresponding neutrino interaction
 cross-section and $P_{osc}$ is the oscillation probability.

 \noindent Furthermore, the $\nu$ intensity can be precisely determined from the
 measurement of the monochromatic $\mu$ current circulating
 in the storage ring (absolute normalization at $1 \%$ level).
 An accurate determination of $\mu$ momentum
 allows for the measurement of the  neutrino energy spectra
 at the detector site.

\noindent The {\large $\nu$F}  lends itself naturally to the exploration of neutrino oscillations
between $\nu$ flavors with high sensitivity to small mixing angles and
small mass differences. The detector should be able to perform both
appearance and disappearance experiments, providing lepton
identification and charge discrimination which is a tag for the initial
flavor and of the oscillation.
   In particular the search  for $\nu_e \rightarrow  \nu_{\mu}$ transitions (``golden channel")
  appears to be very attractive at {\large $\nu$F},  because
  this transition can be studied in appearance  mode looking for
  $\mu^-$ (appearance of wrong-sign $\mu$)
  in  neutrino beams where the neutrino type that is searched for
  is totally absent ($\mu^+$ beam in {\large $\nu$F}).
  With a 40 Kt magnetic detector (MINOS like) exposed to both polarity
  beams and $10^{21}$ muon decays, it will be possible to explore
  the $\theta_{13}$ angle down to $0.1^\circ$ opening the possibility
  to measure the $\delCP$ phase if
  $|\Delta m_{12}^2| \geq 5 {\cdot} 10^{-4}$ eV$^2$ (systematic errors not accounted for)
  \cite{Cervera:2000vy,PilarNufact}.

\noindent Unfortunately, as discussed in Section~\ref{section:numunue},
 the determination of ($\theta_{13},\delCP$)
is not free of ambiguities and up to
eight different regions of the parameter space can fit the same
experimental result.

\begin{figure}
\begin{center}
  \mbox{\epsfig{file=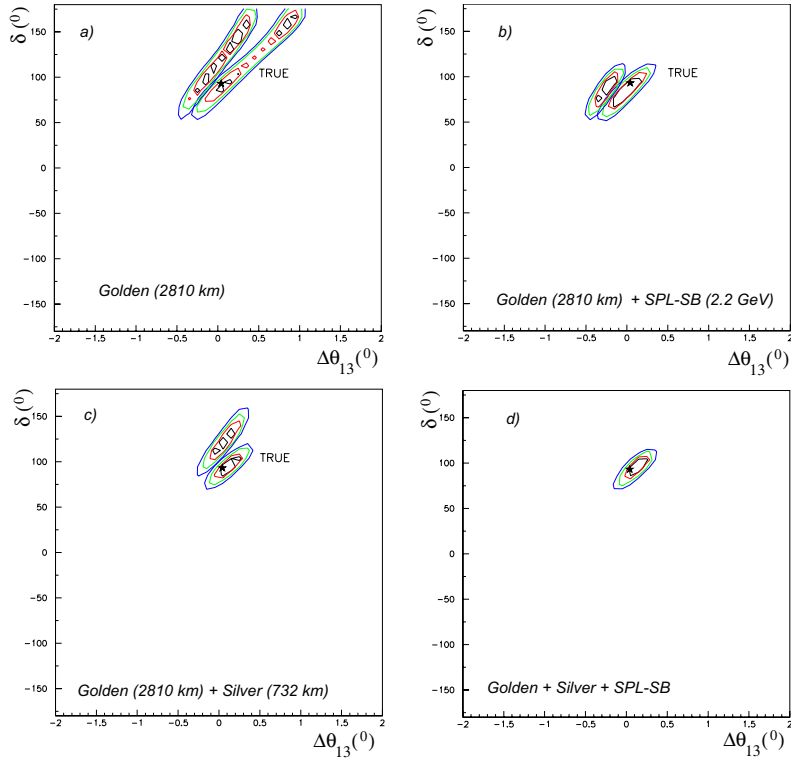,width=0.7\textwidth}}
\end{center}
    \caption{The results of a $\chi^2$ fit for $\bar \theta_{13} =
    2^\circ; \bar \delCP = 90^ \circ$.  Four different combinations of
    experimental data are presented: a) magnetized iron detector (MID) at
    a  $\nu$N; b) MID  plus SPL-SB;
     c) MID plus hybrid emulsion (HE) at $\nu$F; d) the three detectors
    together.  Notice how in case (d) the eightfold-degeneracy is solved
    and a good reconstruction of the physical $\theta_{13},\delCP$ values
    is achieved.}
    \label{super}
\end{figure}

\noindent In order to solve these ambiguities, a single experiment on a single
neutrino beam is not enough. An optimal combination of $\beta$-beams,
SuperBeams and Neutrino Factories has to be considered to deal with
the eightfold degeneracy. Several investigations on how to solve this
problem have been carried out, as reported in~\cite{Donini:2003kr} and
references therein. As an example the result of such an  analysis combining
the golden and the silver ($\nu_e\rightarrow\nu_\tau$) $\nu$F channels
with the SPL-SB, taken from reference
\cite{Autiero:2003fu}, is shown if Fig.~\ref{super}.

More details on the physics performances of a {\large $\nu$F}
 towards a precision measurement of neutrino oscillation parameters
 can be found in Refs.~\cite{APS, nufact2}.

 \section{Conclusions}

 Neutrino oscillations are certainly one of the most important discoveries
 of the recent years.
 They allow to measure fundamental parameters of the Standard
 Model, provide the first insight beyond the electroweak scale,
 play a decisive role in many fields of astrophysics and cosmology and
 allow to explore CP violation in the leptonic sector.

 \noindent The precise measurements of the oscillation parameters and of the still
 unknown parameters
 $\theta_{13}$, \sigdm and \delCP  in the subleading $\nu_\mu \rightarrow \nu_e$
 oscillations in particular
 appear as a must in high energy physics.
  
 Due to the intrinsic difficulties and complexity of the three flavor
 neutrino oscillations
 a single world facility will not probably be sufficient to measure in a firm way all
 these parameters.

 The present generation of neutrino beams at accelerators like NuMI and CNGS can start
 the exploration of the $\theta_{13}$ angle beyond the Chooz limit. 
 However, power and purity
 of these conventional neutrino beams, where neutrinos are generated mainly 
 by  pion and kaon decays
 in a wide range of momenta, seem to limit intrinsically the experimental
 sensitivity.

 New high intensity proton accelerator facilities (in the MWatt regime) are required
 to produce neutrino beams  with an intensity and purity  much higher than  the conventional
 neutrino beams.  
 Novel concept neutrino beams like BetaBeams
 and Neutrino Factories, where neutrinos are produced in the decay of radiocative ions 
 and muons respectively, suitable accelerated to a selected momentum, can open new interesting
 and promising scenarios given  the possibility to explore the neutrino
 oscillation world with high accuracy.
  These new facilities are under study. They will require a long R\&D phase
 with different time-scales due to the different intrinsic difficulties involved in the
 projects and in the constructions. 
 The most physics-ambitious road-map is still to be clarified also because
 no predictions exist for the \thetaot parameter, below the Chooz limit
 $\thetaot<11^\circ$, that drives all the new
 phenomena. 
 
 In the next years SuperBeam facilities, 
 where conventional neutrino beams are improved as flux and purity and tuned to
 $\nu_\mu \rightarrow \nu_e$ transitions, appear the most suitable.

 \noindent The T2K experiment in Japan at the J-Parc accelerator complex,
 using Super-Kamiokande as far detector,
 appears as a reference point in the
 search, well corroborated by the possibility of a second phase
  with a more intense beam and a larger detector of about one megaton mass.

 \noindent The option CERN to Gran Sasso neutrino SuperBeam based on improved synchrotron
 seems to be equivalent to J-Parc II, as far as concerns neutrino fluxes,
 except for $\nu_e$ level
 due to the Off-Axis alignment of J-Parc.  However, it is mandatory to complement it
 with a detector in excess of 100 kton mass to exploit its physics potential
 well beyond the T2K sensitivity.

 \noindent The SPL-SuperBeam project at CERN, complemented with a megaton water Cerenkov detector,
 seems to require a too big effort compared with its physics output, even if the
 3.5 GeV SPL option greatly improves its discovery potential.
 A gigantic water Cerenkov detector would be better exploited if fired by
 a BetaBeam, and it should be stressed that a combination of the SPL-SB with
 a BetaBeam results to be the most powerful facility proposed before the
 Neutrino Factories era.

Smaller and denser detectors could be only used with BetaBeams of higher
energies than that obtainable with the CERN SPS or, in a longer timescale,
 with the Neutrino Factories. Here, iron calorimeters or liquid argon detectors exceeding
 40 Kton of mass, both magnetized in the case of Neutrino Factories, could reach excellent
sensitivities  in neutrino oscillation physics.

\eject


\begin{thebibliography}{999}

 \bibitem{See-Saw}
A.~De Gouvea,
Mod.\ Phys.\ Lett.\ A {\bf 19} (2004) 2799
[arXiv:hep-ph/0503086].

 \bibitem{MSW}
L. Wolfenstein,
Phys. Rev. D17 (1978) 2369.
S.P. Mikheev and A.Y. Smirnov,
Nuovo Cim. C9 (1986) 17.

 \bibitem{Matter_Effects}
 K.~Kimura, A.~Takamura and H.~Yokomakura,
  Phys.\ Rev.\ D {\bf 66}  (2002) 073005
  [arXiv:hep-ph/0205295].
 E.~K.~Akhmedov, R.~Johansson, M.~Lindner, T.~Ohlsson and T.~Schwetz,
  JHEP {\bf 0404} (2004) 078
  [arXiv:hep-ph/0402175].
M.~Freund,
Phys.\ Rev.\ D {\bf 64} (2001) 053003
[arXiv:hep-ph/0103300].

\bibitem{SK-Atmo}
Y.~Fukuda {\it et al.}  [Super-Kamiokande Collaboration],
Phys.\ Rev.\ Lett.\  {\bf 81} (1998) 1562
[arXiv:hep-ex/9807003].

\bibitem{Soudan2}
M.~C.~Sanchez {\it et al.}  [Soudan 2 Collaboration],
Phys.\ Rev.\ D {\bf 68} (2003) 113004
[arXiv:hep-ex/0307069].

\bibitem{Macro}
M.~Ambrosio {\it et al.}  [MACRO Collaboration],
Phys.\ Lett.\ B {\bf 566} (2003) 35
[arXiv:hep-ex/0304037].


\bibitem{SK-L/E}
Y.~Ashie {\it et al.}  [Super-Kamiokande Collaboration],
Phys.\ Rev.\ Lett.\  {\bf 93} (2004) 101801
[arXiv:hep-ex/0404034].


\bibitem{SK-nutau}
S.~Fukuda {\it et al.}  [Super-Kamiokande Collaboration],
Phys.\ Rev.\ Lett.\  {\bf 85} (2000) 3999
[arXiv:hep-ex/0009001].


\bibitem{SK-Final-atm}
Y.~Ashie {\it et al.} [Super-Kamiokande Collaboration],
arXiv:hep-ex/0501064.

\bibitem{Pakvasa}
S.~Pakvasa,
  Pramana {\bf 62} (2004) 347.


\bibitem{K2K}
E.~Aliu {\it et al.}  [K2K Collaboration],
Phys.\ Rev.\ Lett.\  {\bf 94} (2005) 081802
[arXiv:hep-ex/0411038].

\bibitem{Homestake}
B.~T.~Cleveland {\it et al.},
Astrophys.\ J.\  {\bf 496} (1998) 505.

\bibitem{Gallex-GNO}
M.~Altmann {\it et al.}  [GNO Collaboration],
Phys.\ Lett.\ B {\bf 490} (2000) 16
[arXiv:hep-ex/0006034].


\bibitem{Sage}
J.~N.~Abdurashitov {\it et al.}  [SAGE Collaboration],
J.\ Exp.\ Theor.\ Phys.\  {\bf 95} (2002) 181
[Zh.\ Eksp.\ Teor.\ Fiz.\  {\bf 122} (2002) 211]
[arXiv:astro-ph/0204245].


\bibitem{SK-solari}
S.~Fukuda {\it et al.}  [Super-Kamiokande Collaboration],
Phys.\ Lett.\ B {\bf 539} (2002) 179
[arXiv:hep-ex/0205075].

\bibitem{SK-day-night}
M.~B.~Smy {\it et al.}  [Super-Kamiokande Collaboration],
Phys.\ Rev.\ D {\bf 69} (2004) 011104
[arXiv:hep-ex/0309011].

\bibitem{SNO}
S.~N.~Ahmed {\it et al.}  [SNO Collaboration],
Phys.\ Rev.\ Lett.\  {\bf 92} (2004) 181301
[arXiv:nucl-ex/0309004].


\bibitem{Kamland-First}
K.~Eguchi {\it et al.}  [KamLAND Collaboration],
Phys.\ Rev.\ Lett.\  {\bf 90} (2003) 021802
[arXiv:hep-ex/0212021].

\bibitem{Kamland-Final}
T.~Araki {\it et al.}  [KamLAND Collaboration],
Phys.\ Rev.\ Lett.\  {\bf 94} (2005) 081801
[arXiv:hep-ex/0406035].


\bibitem{LSND}
A.~Aguilar {\it et al.}  [LSND Collaboration],
Phys.\ Rev.\ D {\bf 64} (2001) 112007
[arXiv:hep-ex/0104049].

\bibitem{Karmen}
B.~Armbruster {\it et al.}  [KARMEN Collaboration],
Phys.\ Rev.\ D {\bf 65} (2002) 112001
[arXiv:hep-ex/0203021].

\bibitem{Nomad}
P.~Astier {\it et al.}  [NOMAD Collaboration],
Phys.\ Lett.\ B {\bf 570} (2003) 19
[arXiv:hep-ex/0306037].

\bibitem{miniboone}
E.~Church {\it et al.}  [BooNe Collaboration],
arXiv:nucl-ex/9706011.

\bibitem{Chooz} M.~Apollonio {\it et al.} [CHOOZ Collaboration],
 Eur.\ Phys.\ J.\ C {\bf 27} (2003) 331, [arXiv:hep-ex/0301017].

\bibitem{Minos}
  E.~Ables {\it et al.}  [MINOS Collaboration],
Fermilab-proposal-0875
     G.~S.~Tzanakos  [MINOS Collaboration], AIP Conf.\ Proc.\  {\bf 721}, 179 (2004).

 \bibitem{NUMI} The Fermilab NuMI Group, ``NumI Facility Technical Design Report'',
Fermilab Report NuMI-346, 1998.




\bibitem{ICARUS} F.~Arneodo {\it et al.} [ICARUS Collaboration],
Nucl. Instrum. and Meth. A {\bf 461} (2001) 324; P.~Aprili {\it et
al.}, ``The ICARUS experiment'', CERN-SPSC/2002-27, SPSC-P-323.

\bibitem{OPERA} OPERA  Collaboration, CERN-SPSC-P-318, LNGS-P25-00; H.~Pessard  [OPERA Collaboration], arXiv:hep-ex/0504033.
  M.~Guler {\it et al.}  [OPERA Collaboration], ``OPERA: An
appearance experiment to search for $\nu_\mu \rightarrow \nu_\tau$
oscillations in the CNGS beam. Experimental proposal,''
CERN-SPSC-2000-028.

\bibitem{CNGS}    G. Acquistapace {\it et al.}, ``The CERN neutrino beam to
                   Gran Sasso'', CERN 98-02, INFN/AE-98/05 (1998);
                   CERN-SL/99-034(DI), INFN/AE-99/05 Addendum.


 \bibitem{NA20} H.~W.~Atherton {\it et al.}, ``Precise measurements of particle
production by 400 GeV/c protons on Beryllium targets,'' CERN-80-07.



\bibitem{SPY}
G.~Ambrosini {\it et al.}  [NA56/SPY Collaboration],
Eur.\ Phys.\ J.\ C {\bf 10} (1999) 605.

\bibitem{WANF} L.~Casagrande {\it et al.},
``The alignment of the CERN West Area neutrino facility,''
CERN-96-06.



\bibitem{WANF1} A. Guglielmi and G. Collazuol, ``Monte Carlo Simulation of the SPS
                 WANF Neutrino Flux'', INFN/AE-03/05 (2003).\\
P.~Astier {\it et al.}  [NOMAD Collaboration],
Nucl.\ Instrum.\ Meth.\ A {\bf 515} (2003) 800
[arXiv:hep-ex/0306022].

\bibitem{cngs_syst} A.~Ferrari, A.~Guglielmi and P.~Sala, 
to appear in the Proceedings of the NOW 2004 Workshop, Otranto 2004,
[arXiv:hep-ph/0501283].

 \bibitem{FermilabProtons}Report to the Fermilab Director by the Proton
    Committee, November 9, 2004,
   {\tt http://www.fnal.gov/directorate/program\_planning/Nov2004PACPublic/\\Draft\_Proton\_Plan\_v2.pdf}

\bibitem{Komatsu}
M.~Komatsu, P.~Migliozzi and F.~Terranova,
J.\ Phys.\ G {\bf 29} (2003) 443
[arXiv:hep-ph/0210043].

\bibitem{Migliozzi}
P.~Migliozzi and F.~Terranova,
Phys.\ Lett.\ B {\bf 563} (2003) 73
[arXiv:hep-ph/0302274].

\bibitem{SPS_pot_increase}   M.~Benedikt, K.~Cornelis, R.~Garoby, E.~Metral, F.~Ruggiero and
M.~Vretenar, ``Report of the High Intensity Protons Working Group,''
CERN-AB-2004-022-OP-RF.

\bibitem{DChooz} F.~Ardellier {\it et al.} [Double-CHOOZ
Collaboration], arXiv:hep-ex/0405032.

\bibitem{Richter} B. Richter, SLAC-PUB-8587 [arXiv:hep-ph/0008222],
                   and references therein.

\bibitem{PilarNufact}
J.~Burguet-Castell, M.~B.~Gavela, J.~J.~Gomez-Cadenas, P.~Hernandez and O.~Mena,
Nucl.\ Phys.\ B {\bf 608} (2001) 301
[arXiv:hep-ph/0103258].

\bibitem{Minakata}
H.~Minakata and H.~Nunokawa,
JHEP {\bf 0110} (2001) 001
[arXiv:hep-ph/0108085].

\bibitem{Barger:2001yr}
  V.~Barger, D.~Marfatia and K.~Whisnant,
  Phys.\ Rev.\ D {\bf 65} (2002) 073023
  [arXiv:hep-ph/0112119].

\bibitem{Fogli}
G.~L.~Fogli and E.~Lisi,
Phys.\ Rev.\ D {\bf 54} (1996) 3667
[arXiv:hep-ph/9604415].

\bibitem{Harp}
M.G. Catanesi {\it et al.} [HARP Collaboration], CERN-SPSC/2001-017, SPSC/P322, May 2001.

\bibitem{Nufact}
S.~Geer,
Phys.\ Rev.\ D {\bf 57} (1998) 6989
[Erratum-ibid.\ D {\bf 59} (1999) 039903],
[hep-ph/9712290].

\bibitem{nufact2}
 A.~Blondel {\it et al.},
  ``ECFA/CERN studies of a European neutrino factory complex'',
CERN-2004-002.

\bibitem{BetaBeam}
P.~Zucchelli,
Phys.\ Lett.\ B {\bf 532}  (2002) 166.

\bibitem{Fukuda:2002uc}
Y.~Fukuda {\it et al.},
Nucl.\ Instrum.\ Meth.\ A {\bf 501} (2003) 418.

\bibitem{Boger:1999bb} J.~Boger {\it et al.}  [SNO Collaboration],
  Nucl.\ Instrum.\ Meth.\ A {\bf 449}  (2000) 172
  [arXiv:nucl-ex/9910016].


\bibitem{UNO}
C.~K.~Jung [UNO Collaboration]
arXiv:hep-ex/0005046.

\bibitem{T2K} Y. Itow {\it et al.}, ``The JHF-Kamiokande neutrino project'',
                  arXiv:hep-ex/0106019.

\bibitem{hpwf}
A.~C.~Benvenuti {\it et al.},
Phys.\ Rev.\ Lett.\  {\bf 34} (1975) 419.

\bibitem{Monolith}
  F.~Terranova  [MONOLITH Collaboration],
  Int.\ J.\ Mod.\ Phys.\ A {\bf 16S1B} (2001) 736.

\bibitem{Cervera:2000kp}
A.~Cervera, A.~Donini, M.~B.~Gavela, J.~J.~Gomez Cadenas, P.~Hernandez, O.~Mena and S.~Rigolin,
Nucl.\ Phys.\ B {\bf 579} (2000) 17
[Erratum-ibid.\ B {\bf 593} (2001) 731]
[arXiv:hep-ph/0002108].



\bibitem{Cervera:2000vy}
  A.~Cervera, F.~Dydak and J.~Gomez Cadenas,
  Nucl.\ Instrum.\ Meth.\ A {\bf 451}  (2000) 123.

\bibitem{Nova}
  D.~S.~Ayres {\it et al.}  [NOvA Collaboration],
  arXiv:hep-ex/0503053.


\bibitem{Kodama:2000mp}
  K.~Kodama {\it et al.}  [DONUT Collaboration],
  Phys.\ Lett.\ B {\bf 504}  (2001) 218
  [arXiv:hep-ex/0012035].


\bibitem{Silver}
A.~Donini, D.~Meloni and P.~Migliozzi, Nucl.\ Phys.\ B {\bf
646} (2002) 321 [arXiv:hep-ph/0206034].

\bibitem{Autiero:2003fu}
  D.~Autiero {\it et al.},
  Eur.\ Phys.\ J.\ C {\bf 33} (2004) 243
  [arXiv:hep-ph/0305185].

\bibitem{C.Rubbia}
C.~Rubbia, ``The liquid Argon Time Projection Chamber: a new concept for Neutrino
Detector, CERN-EP/77-08.


\bibitem{Icarus-muone}
S. Amoruso {\it et al.} [ICARUS Collaboration],
  Eur.\ Phys.\ J.\ C {\bf 33} (2004) 233
  [arXiv:hep-ex/0311040].

\bibitem{600ton}
Amerio, S. et al. [ICARUS Collaboration], Nucl. Instrum. Meth. A527 (2004) 329-410.

 \bibitem{ARubbia} A. Rubbia: ``Neutrino detectors for future experiments'', invited talk
                   at HIF04 Conference, La Biodola, June 2004 to appear in Nucl. Phys. B Proc.
                   Suppl., arXiv:hep-ph/0412230, and references therein.


\bibitem{targetry}
H.D.~Haseroth {\it et al.}, AIP Proceedings {\bf 721} (2003) 48;
M.S.~Zisman, AIP Proceedings {\bf 721} (2003) 60.

\bibitem{APS}
  C.~Albright {\it et al.}  [Neutrino Factory/Muon Collider Collaboration],
  arXiv:physics/0411123.

\bibitem{Eurisol} {\tt http://www.ganil.fr/eurisol/}

\bibitem{OffAxis}
The E889 Collaboration, "Long Baseline Neutrino Oscillation
Experiment at the AGS", Brookhaven National Laboratory Report BNL
No. 52459, April 1995.
A.~Para and M.~Szleper,
arXiv:hep-ex/0110032.

\bibitem{Takashi}
T.~Kobayashi,
J.\ Phys.\ G {\bf 29} (2003) 1493.
\bibitem{Diwan} M.~V.~Diwan {\it et al.}, Phys.\ Rev.\ D {\bf 68}  (2003) 012002
[arXiv:hep-ph/0303081].

\bibitem{SBall}
H.~Minakata and H.~Sugiyama,
Phys.\ Lett.\ B {\bf 580} (2004) 216
[arXiv:hep-ph/0309323].\\
 P.~Huber, M.~Lindner, M.~Rolinec, T.~Schwetz and W.~Winter,
  Phys.\ Rev.\ D {\bf 70} (2004) 073014
  [arXiv:hep-ph/0403068].
P.~Huber, M.~Lindner, M.~Rolinec, T.~Schwetz and W.~Winter,
  arXiv:hep-ph/0412133.
P.~Huber, M.~Maltoni and T.~Schwetz,
  Phys.\ Rev.\ D {\bf 71}  (2005) 053006
  [arXiv:hep-ph/0501037].
V.~Barger, D.~Marfatia and K.~Whisnant,
Phys.\ Lett.\ B {\bf 560} (2003) 75
[arXiv:hep-ph/0210428].


 \bibitem{CERN-LOW} A. Rubbia and P. Sala, JHEP 209 (2002) 4   [arXiv:hep-ph/0207084].

\bibitem{CNGT}
A.E.~Ball {\it et al.},
``C2GT: intercepting CERN neutrinos to Gran Sasso in the
Gulf of Taranto to measure $\theta_{13}$'',
CERN-SPSC-2004-025, SPSC-M-723.

 \bibitem{CarloRubbia} A. Ferrari {\it et al.}, New J. Phys. {\bf 4} (2002) 88.


\bibitem{SPL-Design}
B.~Autin {\it et al.},
``Conceptual design of the SPL, a high-power superconducting H- linac  at CERN,''
CERN-2000-012.

 \bibitem{SPL-Physics}
J.~J.~Gomez-Cadenas {\it et al.},
 Proceedings of ``Venice 2001, Neutrino telescopes'', vol. 2*, 463-481,
arXiv:hep-ph/0105297.
A.~Blondel {\it et al.},
Nucl.\ Instrum.\ Meth.\ A {\bf 503} (2001) 173.
M. Mezzetto,
 J. Phys. G {\bf 29} (2003) 1771 [arXiv:hep-ex/0302005].

 \bibitem{nufact1}
  M.~Apollonio {\it et al.},
  arXiv:hep-ph/0210192.


\bibitem{Cazes}
J.~E.~Campagne and A.~Cazes,
arXiv:hep-ex/0411062.

\bibitem{Garoby-SPL}R.~Garoby, ``The SPL at CERN,'' CERN-AB-2005-007.

\bibitem{MMNufact04}
M. Mezzetto, to be published in Proceedings of Nufact04.

\bibitem{Lindroos} B.~Autin {\it et al.}, arXiv:physics/0306106.
M. Benedikt, S. Hancock and M. Lindroos,
Proceedings of EPAC 2004,
http://accelconf.web.cern.ch/AccelConf/e04.

\bibitem{beta} M. Mezzetto, J.Phys. G {\bf 29 } (2003) 1781 [arXiv:hep-ex/0302007].
J.~Bouchez, M.~Lindroos and M.~Mezzetto,
AIP conference proceedings, {\bf 721} (2003)  37  [arXiv:hep-ex/0310059].
M.~Mezzetto,
  Nucl.\ Phys.\ Proc.\ Suppl.\  {\bf 143} (2005) 309
  [arXiv:hep-ex/0410083].

\bibitem{Donini:2004hu}
A.~Donini, E.~Fernandez-Martinez, P.~Migliozzi, S.~Rigolin and
L.~Scotto Lavina, Nucl.\ Phys.\ B {\bf 710} (2005) 402
[arXiv:hep-ph/0406132].


\bibitem{MatsPrivate}
M. Lindroos, EURISOL DS/TASK12/TN-05-02.


\bibitem{latestJJ}
J.~Burguet-Castell, D.~Casper, E.~Couce, J.~J.~Gomez-Cadenas and P.~Hernandez,
  arXiv:hep-ph/0503021.



\bibitem{HighEnergy}
J.~Burguet-Castell {\it et al.},
Nucl.\ Phys.\ B {\bf 695} (2004) 217
[arXiv:hep-ph/0312068].

\bibitem{HighEnergy2}
F.~Terranova, A.~Marotta, P.~Migliozzi and M.~Spinetti,
Eur.\ Phys.\ J.\ C {\bf 38} (2004) 69
[arXiv:hep-ph/0405081].

\bibitem{SuperSPS}
 O.~Bruning {\it et al.},
``LHC luminosity and energy upgrade: A feasibility study,''
CERN-LHC-PROJECT-REPORT-626.


\bibitem{Donini:2003kr}
  A.~Donini,
  AIP Conf.\ Proc.\  {\bf 721} (2004) 219
  [arXiv:hep-ph/0310014].



\end{thebibliography}
  \end{document}